\documentclass{aa}  

\usepackage{graphicx}
\usepackage{txfonts}
\usepackage{hyperref}

\usepackage{xcolor}
\usepackage{orcidlink}

\definecolor{Blue}{rgb}{0,0.08,0.65}
\definecolor{Blue2}{rgb}{0,0.4,0.6}

\hypersetup{
    colorlinks,
    linkcolor={red!60!black},
    citecolor={Blue2},
    urlcolor={Blue}
}

\newcommand{\msun}{\mbox{$\rm{M}_\odot$}\xspace}

\newcommand{\hcc}{\mbox{$\rm{H} \, \rm{cm}^{-3}$}\xspace}

\newcommand{\pymc}{\mbox{{\sc \small PyMC3}}\xspace}
\newcommand{\ssfr}{\mbox{sSFR}\xspace}
\newcommand{\sfr}{\mbox{SFR}\xspace}

\newcommand{\mstar}{\mbox{$M_\star$}\xspace}
\newcommand{\hagn}{\mbox{{\sc \small Horizon-AGN}}\xspace}
\newcommand{\newh}{\mbox{{\sc \small NewHorizon}}\xspace}
\newcommand{\vintergatan}{\mbox{{\sc \small Vintergatan}}\xspace}
\newcommand{\ramses}{\mbox{{\sc \small Ramses}}\xspace}
\newcommand{\hi}{{\sc H\,i}\xspace}
\newcommand{\hmol}{{\sc H$_2$}\xspace}
\newcommand{\sigmagas}{\mbox{$\Sigma_{\rm gas}$}\xspace}
\newcommand{\sigmagasSF}{\mbox{$\Sigma_{\rm H_2}$}\xspace}
\newcommand{\sigmagastot}{\mbox{$\Sigma_{\rm \textsc{H\,i} + \rm{H}_2}$}\xspace}
\newcommand{\sigmasfr}{\mbox{$\Sigma_{\rm SFR}$}\xspace}

\newcommand{\mgassf}{\mbox{$M_{\rm H_2}$}\xspace}

\newcommand{\mach}{\mbox{$\mathcal{M}$}\xspace}

\newcommand{\effr}{\mbox{$R_{\rm eff}$}\xspace}

\begin{document}

   \title{Emergence and cosmic evolution of the Kennicutt-Schmidt relation driven by interstellar turbulence}
   \titlerunning{Kennicutt-Schmidt relation}
   \authorrunning{K. Kraljic et al.}
  
   \author{Katarina~Kraljic\inst{1}\orcidlink{0000-0001-6180-0245} \and
          Florent~Renaud\inst{2,3}\orcidlink{0000-0001-5073-2267} \and
          Yohan~Dubois\inst{4}\orcidlink{0000-0003-0225-6387} \and
          Christophe~Pichon\inst{4,5,6} \and
          Oscar~Agertz\inst{2}\orcidlink{0000-0002-4287-1088} \and
          Eric~Andersson\inst{7}\orcidlink{0000-0003-3479-4606}\and
          Julien~Devriendt\inst{8} \and
          Jonathan~Freundlich\inst{1}\orcidlink{0000-0002-5245-7796} \and
          Sugata~Kaviraj\inst{9}\orcidlink{0000-0002-5601-575X} \and
          Taysun~Kimm\inst{10} \and
          Garreth~Martin\inst{11,12} \and
          S\'ebastien~Peirani\inst{13} \and
          \'Alvaro~Segovia~Otero\inst{2}\orcidlink{0000-0002-0415-3077} \and
          Marta~Volonteri\inst{4} \and
          Sukyoung~K.~Yi\inst{10}\orcidlink{0000-0002-4556-2619}
          }

   \institute{Observatoire Astronomique de Strasbourg, Universit\'e de Strasbourg, CNRS, UMR 7550, F-67000 Strasbourg, France\\
              \email{katarina.kraljic@astro.unistra.fr}
         \and
             Lund Observatory, Division of Astrophysics, Department of Physics, Lund University, Box 43, SE-221 00 Lund, Sweden
         \and
             University of Strasbourg Institute for Advanced Study, 5 all\'ee du G\'en\'eral Rouvillois, F-67083 Strasbourg, France
         \and
             Institut d'Astrophysique de Paris, CNRS and Sorbonne Universit\'e, UMR 7095, 98 bis Boulevard Arago, F-75014 Paris, France
        \and
             IPhT, DRF-INP, UMR 3680, CEA, L'Orme des Merisiers, B\^at 774, 91191 Gif-sur-Yvette, France
        \and
             Korea Institute for Advanced Study, 85 Hoegi-ro, Dongdaemun-gu, Seoul 02455, Republic of Korea
        \and 
             Department of Astrophysics, American Museum of Natural History, 200 Central Park West, New York, NY 10024, USA
        \and
            Department of Physics, University of Oxford, Keble Road, Oxford OX1 3RH, United Kingdom     
        \and
            Centre for Astrophysics Research, Department of Physics, Astronomy and Mathematics, University of Hertfordshire, Hatfield AL10 9AB, UK
        \and 
            Department of Astronomy and Yonsei University Observatory, Yonsei University, Seoul 03722, Republic of Korea
        \and 
            Korea Astronomy and Space Science Institute, 776 Daedeokdae-ro, Yuseong-gu, Daejeon 34055, Republic of Korea
        \and 
            Steward Observatory, University of Arizona, 933 N. Cherry Ave, Tucson, AZ 85719, USA
        \and
            Observatoire de la C\^ote d'Azur, CNRS, Laboratoire Lagrange, Bd de l'Observatoire, CS 34229, F-06304 Nice Cedex 4, France
             }

   \date{Received July 1, 2023; accepted }

\abstract{The scaling relations between the gas content and star formation rate of galaxies provide useful insights into processes governing their formation and evolution. We investigate the emergence and the physical drivers of the global Kennicutt-Schmidt (KS) relation at $0.25 \leq z \leq 4$ in the cosmological hydrodynamic simulation \newh capturing the evolution of a few hundred galaxies with a resolution of $\sim$ 40 pc. The details of this relation vary strongly with the stellar mass of galaxies and the redshift. A power-law relation $\sigmasfr \propto \Sigma_{\rm gas}^{a}$ with $a\approx 1.4$, like that found empirically, emerges at $z\approx 2\-- 3$ for the most massive half of the galaxy population. However, no such convergence is found in the lower-mass galaxies, for which the relation  gets shallower with decreasing redshift. At the galactic scale, the star formation activity correlates with the level of turbulence of the interstellar medium, quantified by the Mach number, rather than with the gas fraction (neutral or molecular), confirming previous works.
With decreasing redshift, the number of outliers with short depletion times diminishes, reducing the scatter of the KS relation, while the overall population of galaxies shifts toward low densities. Using pc-scale star formation models calibrated with local Universe physics, our results demonstrate that the cosmological evolution of the environmental and intrinsic conditions conspire to converge towards a significant and detectable imprint in galactic-scale observables, in their scaling relations, and in their reduced scatter.}

   \keywords{galaxies: evolution -- galaxies: star formation                    -- galaxies: ISM --
                turbulence --
                methods: numerical
               }
   \maketitle
%

\section{Introduction}

Decades of observational works have highlighted scaling relations between the gas content of galaxies and their star formation rate (\sfr), as a key ingredient of galaxy formation. The pioneering work of \citet{Schmidt1959} revealed a tight relation between the densities of gas and \sfr. It has later been complemented by \citet{Kennicutt1989}, which proposed an empirical relation between the surface densities of neutral gas and \sfr, of the form $\sigmasfr \propto \Sigma_{\rm gas}^a$ with an index $a\approx 1.4$ for local star-forming galaxies. Since, a number of studies have extended the range of this Kennicutt-Schmidt (KS) relation by considering a more diverse population of galaxies like local starbursts \citep[e.g.][]{Kennicutt1998}, high redshift discs \citep[e.g.][]{Tacconi2010}, sub-millimeter galaxies \citep[e.g.][]{Bouche2007}, and sub-galactic scales in local galaxies \citep[e.g.][]{Bigiel2008} to name a few (see \citealt{Kennicutt2012} for a review).

The inferred slope of $\sim 1.4$ for nearby normal spirals \citep{Kennicutt1989}, recently confirmed by \cite{delosReyes2019} in their revisited analysis, is consistent with measurements at higher redshifts \citep[$z=1.5$,][]{Daddi2010}. The slope of $1.4-1.5$
has also been found for the combined sample of normal and starbursting local galaxies \citep{Kennicutt2021}. On the other hand, dwarf galaxies yield slopes closer to unity \citep[e.g.][]{Filho2016, Roychowdhury2017}, such that including them in the samples lowers the slope to $\sim 1.3$ and increases the scatter of the KS relation \citep{delosReyes2019}. Starburst galaxies alone appear to result in different values of a slope for different samples. For instance, \cite{Kennicutt2021} suggests values of $1-1.2$, shallower than previous findings ($\approx 1.3-1.4$, see \citealt{Kennicutt1989,Daddi2010}). 
However, there seems to be a consistency in the findings of evidence for a bimodal (or even multimodal) relation for starbursts and non-starbursting galaxies, yet with significant overlap \citep[e.g.][]{Daddi2010,Genzel2010,Kennicutt2021}. 

As the neutral gas phase also includes diffuse atomic gas, that is yet to collapse, the SFR correlates more strongly with  the molecular gas contents alone \citep[e.g.][]{Bigiel2011}. This motivated the introduction of another KS-like relation, molecular one, with a slope empirically found to be close to unity in nearby star-forming galaxies both on galactic \citep[e.g.][]{Kennicutt1998,Liu2015,delosReyes2019} and sub-kpc scales \citep[e.g.][]{Bigiel2008,Leroy2008,Leroy2013,Onodera2010,Schruba2011,Sun2023}, in local \citep[e.g.][]{Liu2015} and high-redshift starbursts \citep[e.g.][]{Sharon2013,Rawle2014}, and high-redshift galaxies both in galaxy-averaged \citep[e.g.][]{Genzel2010,Tacconi2013,Freundlich2019} and spatially resolved studies \citep[e.g.][]{Freundlich2013,Genzel2013}. 
Considering the molecular gas alone reduces the variations in the measured slopes across these families of galaxies.

This wealth of observational studies comes with a vast diversity of resolutions, scales, tracers, conversion factors, and fitting methods, which make comparisons and compilations delicate \citep[see e.g.][]{delosReyes2019}.
For instance, \cite{Sun2023} found that systematic uncertainties in the estimation of slopes due to different choices of SFR calibrations \citep[see also e.g.][on the impact of the adopted extinction model]{Genzel2013} may be of about 10\% to 15\%, while CO-to-H$_2$ conversion factor may produce an additional 20\% to 25\% \citep[in qualitative agreement with e.g.][]{Liu2015,delosReyes2019}. Despite uncertainties on the values to adopt for such conversion factors \citep{Bolatto2013}, both observations and simulations report non-negligible variations across galactic disks \citep{Teng2023}, and from galaxy to galaxy \citep{Narayanan2011}, caused by the underlying range of the physical conditions. This is particularly important in starbursting galaxies which yield a significantly lower CO-to-H$_2$ conversion factor than the standard Milky Way value \citep{Renaud2019a}. This adds to uncertainties on the slope and scatter of the KS relation of heterogeneous samples.

Similarly, the choice of the fitting method was also found to have a significant effect on the derived slopes \citep[e.g.][]{Shetty2013,Kennicutt2021}. \cite{delosReyes2019} recently revisited the KS relation for non-starbursting galaxies comparing three widely used fitting techniques, ordinary linear regression, bivariate regression, and hierarchical Bayesian \textit{linmix} model, finding changes for the inferred slope of up to $\sim$ 30\%.

Nowadays, this variety of results seems to be acknowledged as being mostly due to systematics related to the above-mentioned methodological choices. Yet, there is still a poor understanding of the physics behind the intrinsic scatter of the KS relation. Ongoing and future missions are opening new windows on the physics of the earlier Universe, in particular on the star formation activity of galaxies during their first few Gyr thanks to the James Webb Space Telescope. To accompany these efforts, cosmological simulations can provide insights into the behaviors of the current models in these high redshift conditions \citep[e.g.][]{Kravtsov2003,Feldmann2012,Semenov2019}. Models and sub-grid prescriptions for star formation and feedback are calibrated using detailed observations in the local Universe, and even mainly from the Solar neighborhood. It is thus important to understand how they behave when applied to different environments. In particular, identifying at which cosmic epoch a given scaling relation emerges is a crucial step in the interpretation of observations at high redshift, and the further confrontation with existing models. 

This first paper of a series, intended to complement our understanding of the evolution of the physics of star formation across cosmic time, focuses on the question of when the KS relation emerges, how it evolves, and what physical parameters are primarily driving it. To address these questions, we use the large-scale zoom-in hydrodynamic simulation \newh \citep[][]{Dubois2021}, and perform an analysis of the KS relation on the galactic scale and at different cosmic epochs, from redshift 4 down to 0.25.

The remainder of this paper is structured as follows. Section~\ref{sec:methods} describes the simulated data set and methods used in the analysis. Section~\ref{sec:KS} presents the results on the emergence of the KS relation and its dependence on different physical properties of galaxies. These results are discussed in Section~\ref{sec:discuss} and finally, Section~\ref{sec:conclusion} concludes.

\section{Methods}
\label{sec:methods}

\begin{figure*}
\centering
\includegraphics[width=0.8\textwidth]{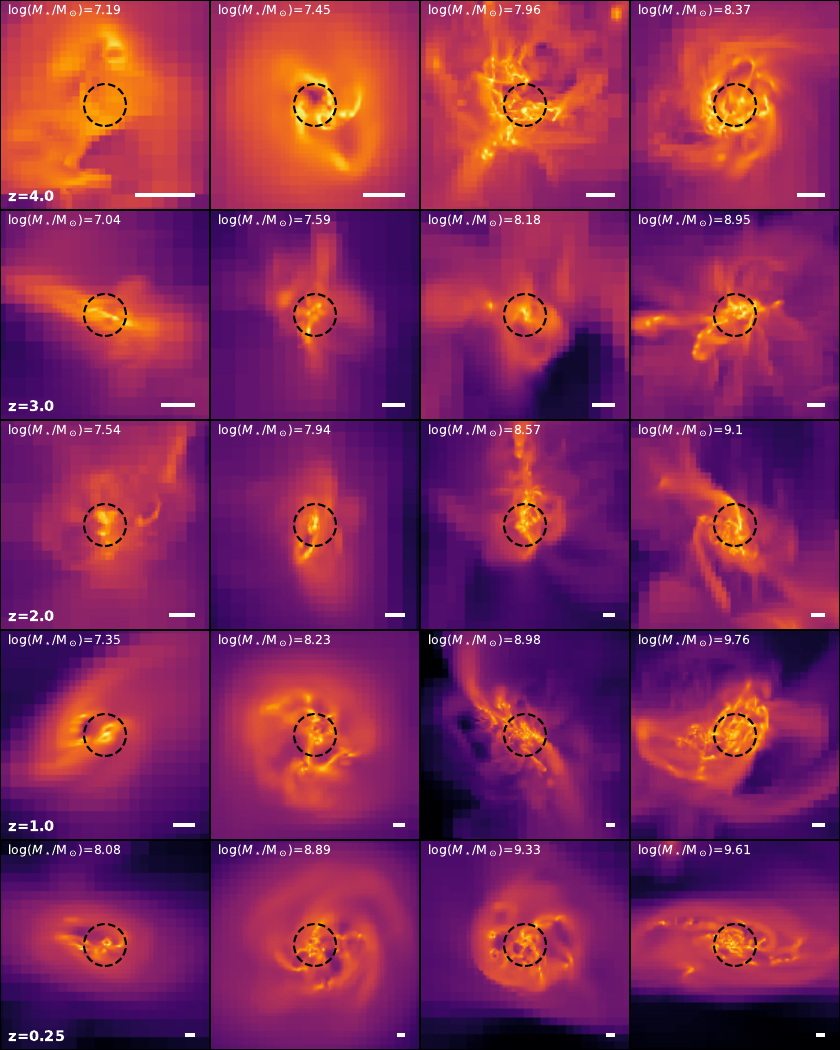}
\caption{Projection of gas density, within the 10$R_{\rm{eff}}$ thick slices, of representative galaxies at different redshifts (rows) and stellar masses (columns). Dashed circles show the effective radius of the stellar component (see Section~\ref{sec:postprocessing} for the definition), and the white horizontal bars indicate a 1 kpc scale.
}
\label{fig:mosaic}
\end{figure*}

\subsection{The NewHorizon simulation}

This work makes use of the \newh\footnote{\href{http://new.horizon-simulation.org}{http://new.horizon-simulation.org}} simulation \citep{Dubois2021}, a large-scale zoom-in simulation of a sub-volume extracted from the large-scale cosmological simulation \hagn \citep{Dubois2014,Kaviraj2017}. Combining a relatively large volume with a resolution typical of standard zoom-in simulations, \newh captures the structure of the cold interstellar medium of several hundreds of galaxies. 
This allows us to fully resolve the wider cosmic environment as well as emergently produce a realistic distribution of galaxy properties.
Therefore, we are able to perform statistical studies on many galaxy properties, at an unprecedented resolution over such volumes.

\newh reproduces reasonably well many observables \citep[see][]{Dubois2021}, such as the galaxy stellar mass function, the cosmic SFR density, the stellar density, the stellar mass-star formation rate main sequence, galaxy gas fractions, the specific SFR-mass relation, the size-mass relation, the mass-metallicity relation, and the Tully-Fisher relation \citep[see also e.g.][]{Volonteri2020,Jackson2021a,Jackson2021b,Martin2021,Park2021,Grisdale2022}. 
The details of the simulation can be found in \cite{Dubois2021}, and we describe here only the features of interest for the analysis of the KS relation.

The \newh simulation is run with the adaptive mesh refinement code \ramses~\citep{teyssier02}, with $\Lambda$CDM cosmology compatible with the WMAP-7 data~\citep{komatsuetal11}. The mass resolution is $1.2\times 10^6 \,\rm M_\odot$ for the dark matter and $1.3\times 10^4 \,\rm M_\odot$ for the stars. The refinement strategy allows to reach the spatial resolution of up to 34 pc.

\newh includes heating of the gas from a uniform UV background following \cite{haardt&madau96} and models the self-shielding of the ultraviolet background in optically thick regions following \cite{rosdahl&blaizot12}. Gas cooling down to $\approx 10^4\, \rm K$ is allowed through collisional ionization, excitation, recombination, Bremsstrahlung, and Compton cooling. Further cooling of metal-enriched gas down to $0.1\, \rm K$ follows tabulated rates from \cite{dalgarno&mccray72} and \cite{sutherland&dopita93}.

Gas above the density threshold of 10 \hcc is converted into stars following the Schmidt relation $\dot \rho_\star= \epsilon_\star {\rho / t_{\rm ff}}$, where $\dot \rho_\star$ is the star formation rate density, $\rho$ the gas mass density, $t_{\rm ff}=\sqrt{3\pi/(32G\rho)}$ the local free-fall time of the gas, $G$ the gravitational constant, and $\epsilon_\star$ is a varying star formation efficiency \citep[see][]{krumholz&mckee05,padoan&nordlund11, hennebelle&chabrier11,federrath&klessen12,kimmetal17,trebitschetal17,trebitschetal20}. 
$\epsilon_\star$ is a function of the local turbulence Mach number ${\cal M}$, and the virial parameter $\alpha_{\rm vir}=2E_{\rm kin}/E_{\rm grav}$ ($E_{\rm kin}$ and $E_{\rm grav}$ are respectively the turbulent and gravitational energies):
\begin{equation}
  \label{eq:sfr_ff}
  \epsilon_{\star} = \epsilon_\star({\cal M},\alpha_{\rm vir}) = \frac{\epsilon}{2\phi_{\rm t}} \exp\left(\frac{3}{8}\sigma_s^2\right)\left[1 + \mathrm{erf}\left(\frac{\sigma_s^2 - s_{\rm crit}}{\sqrt{2\sigma_s^2}}\right)\right],
\end{equation}
where $s_{\rm crit}(\alpha_{\rm vir},\mathcal{M})$ is the critical logarithmic density contrast of the gas density probability distribution function with variance 
$\sigma_{\rm s}^2(\mathcal{M})$ \citep[see][for details]{Dubois2021}.
The parameter $\phi_{\rm t}$ is set to the best-fit value between the theory and the numerical experiments~\citep{federrath&klessen12} and $\epsilon$, set to 0.5, mimics proto-stellar feedback effects to regulate the amount of gas eligible to form stars \citep{matzner&mckee00,alvesetal07,andreetal10}.
In short, this prescription favors the rapid formation of stars in dense, gravitationally collapsing medium with compressible turbulence.

\newh includes feedback from type II supernovae \citep[][]{Thornton1998} following the mechanical supernova feedback scheme of \citet[\citealt{kimmetal15}]{kimm&cen14} to ensure a correct amount of radial momentum transfer. \newh also follows the formation, growth, and dynamics of massive black holes and the associated feedback from active galactic nuclei, following two different modes depending on the Eddington rate~\citep{duboisetal12}. At low accretion rates, the massive black hole powers jets releasing mass, momentum, and total energy into the gas (the so-called radio mode feedback, \citealt{duboisetal10}), while at high rates, it releases only thermal energy (the so-called quasar mode,~\citealp{teyssieretal11}).

\subsection{Postprocessing and sample selection}
\label{sec:postprocessing}

Galaxies are identified with the \textsc{AdaptaHOP} halo finder~\citep{aubert04} run on the stellar particle distribution (see \citealt{Dubois2021} for details).
This work employs the 100\% purity sample, i.e. halos and embedded galaxies devoid of low-resolution DM particles.\footnote{Given that \newh is a zoom simulation embedded in a larger cosmological volume filled with lower DM resolution particles, some halos of the zoom regions can be polluted with low-resolution DM particles.}

Following the convention adopted in \citet{Dubois2021}, we identify the neutral gas component (atomic and molecular), noted \hi~+~ \hmol, as denser than 0.1 \hcc and colder than $2~\times~10^4\,\rm~K$, and the \hmol molecular component denser than 10 \hcc and colder than $2\times10^4\,\rm K$. Reproducing the ionisation and molecular states of the gas would require a detailed treatment of radiative transfer and molecular chemistry, out of the scope of this paper. 
The surface densities of neutral (\sigmagastot) and molecular gas (\sigmagasSF), and of the star formation rate (\sigmasfr) are computed within the (three-dimensional) effective radius $R_{\rm eff}$ of each galaxy, defined as the geometric mean of the half-mass radius of the projected stellar densities along each of the Cartesian axes \citep[see][for more details]{Dubois2021}.

We do not de-project the galaxies when computing surface densities, and note that this is not expected to have a strong impact on statistical distributions of surface densities \citep[e.g.][]{Appleby2020}. 
The star formation rate is estimated by considering only the stars younger than 10 Myr, consistent with the timescale probed by the H$\alpha$-based \sfr indicator. This choice is a compromise, as longer time scales would tend to include the effects of stellar feedback on the properties of the interstellar medium and increase the intrinsic scatter of the $\sigmagas-\sigmasfr$ relation \citep[e.g.][]{Feldmann2012,Andersson2021}, in particular at higher redshifts. 

The turbulence Mach number \mach of each gas cell is computed as $\mach=\sigma_{\rm g}/(\sqrt{3} c_{\rm s})$, where $\sigma_{\rm g}$ and $c_{\rm s}$ are its velocity dispersion sound speed, respectively
\citep[see][for more details]{Kraljic2014}.
Then, the Mach number of the galaxy is given by the mass-weighted average of \mach of every neutral gas (\hi $+$ \hmol) cell\footnote{However, considering the molecular gas alone yields qualitatively similar results (not shown).}
within the effective radius $R_{\rm eff}$.

In this paper, we analyse the population of galaxies at the redshifts 4, 3, 2, 1, and 0.25. We consider only the galaxies with stellar masses $M_\star$ above $10^7\, \msun$, and having hosted star formation in the last 10 Myr. At $z=4$, this corresponds to $\sim$ 90\% of the entire sample of galaxies with $M_\star \geq 10^7\, \msun$, while with decreasing redshift this fraction decreases to $\sim$ 40\% at $z=0.25$. This is essentially due to the lack of star formation activity during the last 10 Myr and is limited to galaxies with  $M_\star \lesssim 10^{8.5}\, \msun$ at $z\geq2$, while below $z\sim2$, more and more massive galaxies are concerned.
Only $\lesssim2\%$ of galaxies at $z=4-1$ and $\sim 7\%$ at $z=0.25$ do not host any neutral gas within their effective radius and these are limited to low mass range ($\lesssim10^{8}$ \msun) at all redshifts.
The resulting numbers of galaxies at each redshift are provided in Table~\ref{tab:numbers}.
Examples of representative galaxies from the various stellar mass bins used in the analysis and at different redshifts are shown in Fig.~\ref{fig:mosaic}. 
The stellar mass bins adopted throughout the paper are defined using the quartiles of the mass distribution at each redshift, and thus yield evolving ranges as the overall population grows.

\begin{table}
\centering
\caption{Number of galaxies and median of their stellar mass (in log \msun) used in our analysis, at each redshift. Note that our sample is limited to galaxies of stellar mass $\geq 10^7$ \msun, with \sfr~$\geq 10^{-3}\,\msun\,\rm yr^{-1}$ and containing neutral gas.}
\begin{tabular}{lcc}
\hline
$z$& Number of galaxies & Median \mstar \\
\hline
4 & 535 & 7.55\\
3 & 558 & 7.74\\
2 & 582 & 7.99\\
1 & 303 & 8.63\\
0.25 & 153 & 9.07\\
\hline
\end{tabular}
\label{tab:numbers}
\end{table}

\subsection{Fitting method}

To quantify the correlation between the surface densities of gas and SFR, we fit the distributions with the relation $\log \Sigma_{\rm SFR}~=~a(\log \Sigma_{\rm gas}) + b$, with the best-fit values for the slope $a$ and intercept $b$. In this paper, we do not attempt to provide a thorough study of the impact of different fitting methods used in the literature on the estimated values for the obtained parameters \citep[we refer the readers to e.g.][for a discussion on fitting methods used in science]{Hogg2010}.
Nevertheless, we compare three different fitting methods: the ordinary least square (OLS) technique, the OLS bisector technique \citep{Isobe1990}, and the Bayesian linear regression. 
The results of the Bayesian regression are shown throughout the paper, as it provides a more robust treatment of errors and is thus particularly adapted to observational measures. In Appendix~\ref{sec:bayesian}, we report the results of the OLS bisector, together with a more detailed comparison of different methods. 
In short, all three fitting methods provide a qualitatively similar trend for the slopes and dispersions around the best fit as a function of redshift and stellar mass. We note however that quantitatively, the values of the slope differ: they are systematically higher for the bisector OLS method, and the dispersion around the best fit is also systematically higher. Overall, depending on the population of galaxies and the gas tracer under consideration, the choice of the fitting method can produce changes of 10\% to 30\% for the slope, in agreement with similar estimates of the impact of fitting algorithms on recent observational data \citep[see][]{delosReyes2019,Kennicutt2021}.
These differences should be kept in mind when comparing the values reported in the literature.

\section{Results}
\label{sec:KS}

\subsection{Distributions of galaxies in the KS plane}
\label{subsec:mass_ssfr}

\begin{figure*}
\centering\includegraphics[width=0.98\textwidth]{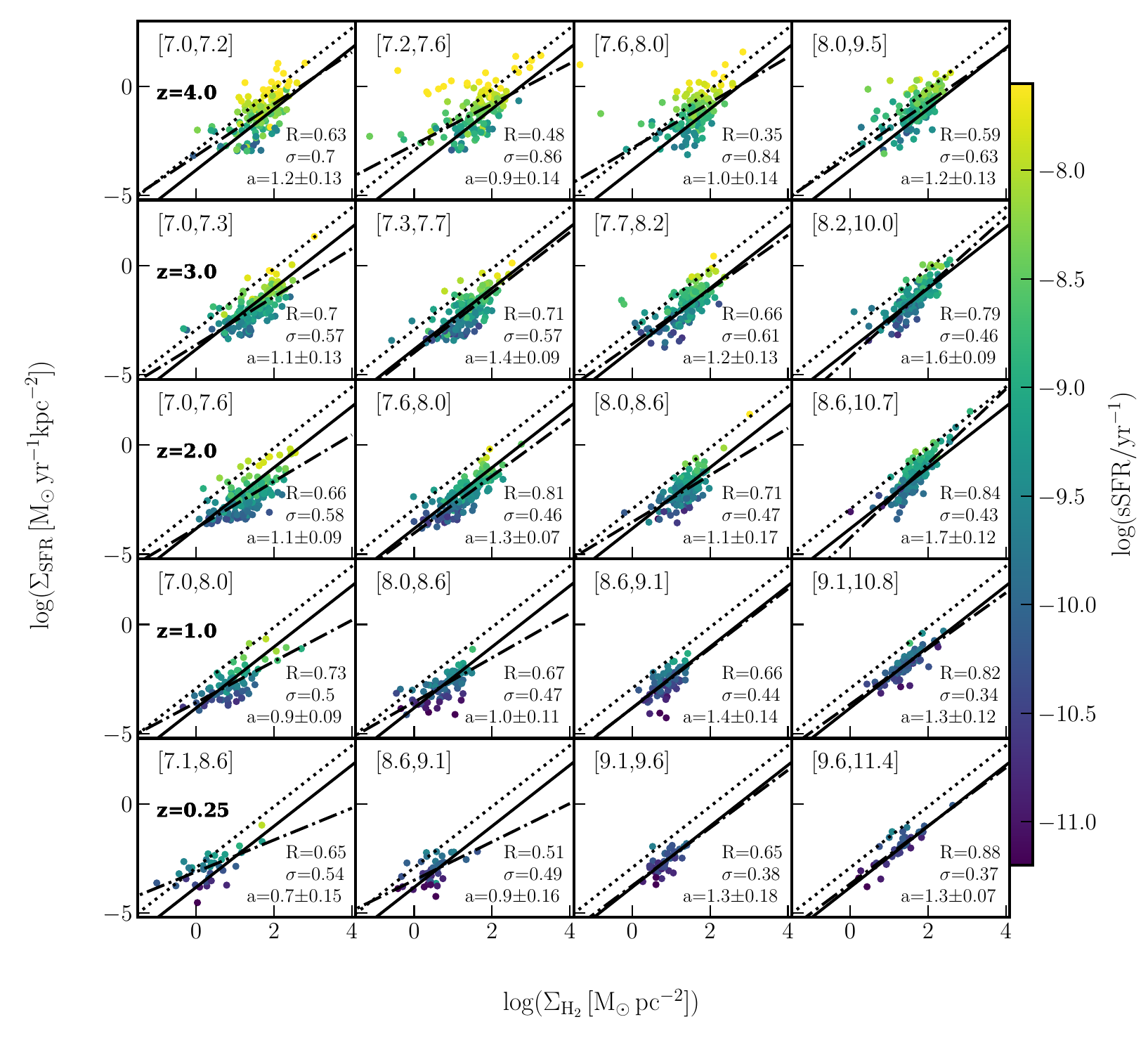}
\caption{Distribution of galaxies in the KS plane at different redshifts (rows), and in four equally-populated stellar mass quartiles at each redshift (columns). The mass range of each quartile is shown in square brackets (in log \msun). The colours indicate the specific star formation rate of the galaxies, measured over a time scale of 10 Myr. Dash-dotted lines are fits at each redshift and mass bin, with the slope $a$ and $\sigma$, the standard deviation of residuals of the best-fit relation, shown in the lower right corners. The coefficient R shown on the bottom right of each panel is the Pearson correlation coefficient. 
The solid black and dotted black lines show the sequence of discs and starbursts, respectively, from \protect \cite{Daddi2010}, for reference. Note, however, that the fitting method differs from the one adopted in this work. Regardless of stellar mass, the distributions of galaxies move within the KS plane towards lower values of \sigmagasSF and \sigmasfr with decreasing redshift. At each redshift and in each stellar mass bin, the \ssfr of galaxies strongly correlates with \sigmasfr. A version of this figure using the neutral gas is available in Fig.~\ref{fig:KS_SFR10_gastot_ssfr}.}
\label{fig:KS_SFR10_gasSF_ssfr}
\end{figure*}

\begin{figure*}
\centering\includegraphics[width=0.9\textwidth]{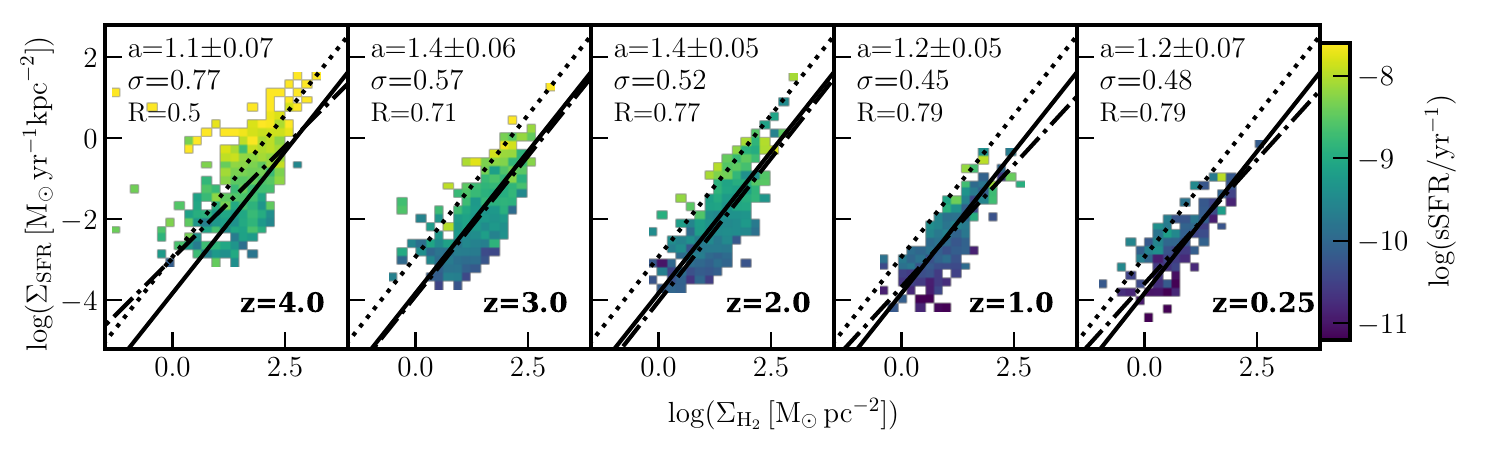}\\[-0.1cm]
\centering\includegraphics[width=0.9\textwidth]{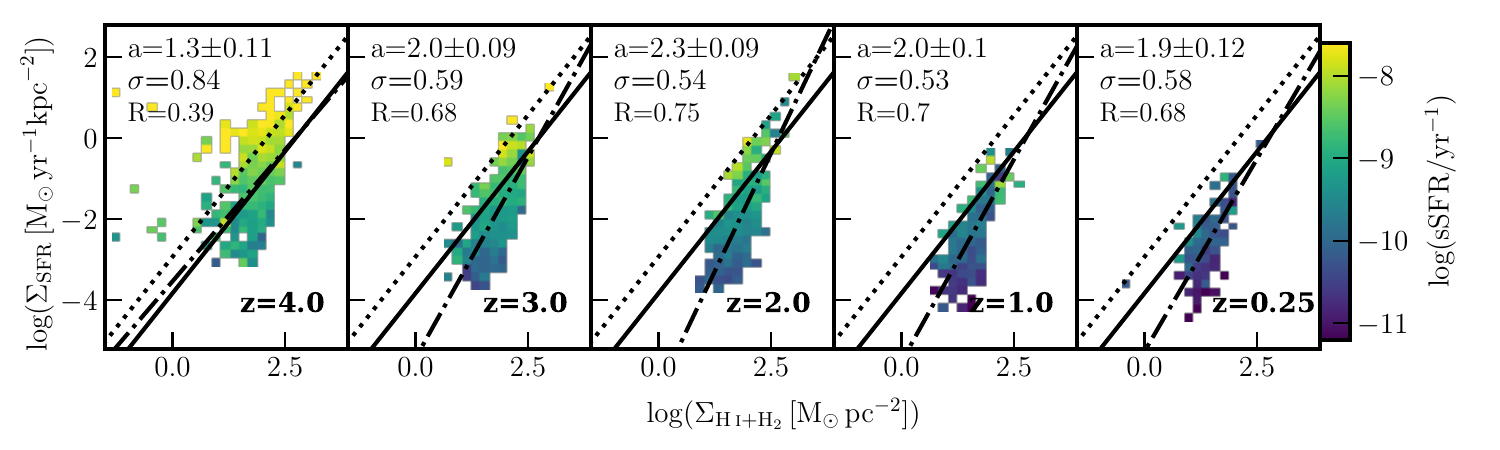}

\caption{Same as Figs.~\ref{fig:KS_SFR10_gasSF_ssfr} and~\ref{fig:KS_SFR10_gastot_ssfr}, but without binning the stellar masses, and considering the molecular gas only (top), and the neutral gas (bottom). The correlation between \ssfr and \sigmasfr seen at in different stellar mass bins (Fig/~\ref{fig:KS_SFR10_gasSF_ssfr}) is still apparent when stacking all galaxies. At all redshifts, the slope ($a$) and the dispersion ($\sigma$) around the best-fit relation (dash-dotted line) are larger for the neutral gas than for the molecular gas. At all redshifts, the correlation is stronger for molecular gas than for neutral gas.}
\label{fig:KS_ssfr_gas_SF_tot}
\end{figure*}

We start by investigating the diversity of star-forming galaxies and its evolution with cosmic time, by analysing the distributions of galaxies in the Kennicutt-Schmidt (KS) plane in different stellar mass bins, and as a function of redshift. 
Figure~\ref{fig:KS_SFR10_gasSF_ssfr} shows the distribution of galaxies of the \newh simulation in the \sigmagasSF-\sigmasfr plane, in different stellar mass bins (columns) at different redshifts (rows), colour-coded by their specific star formation rate (\ssfr~$=~\sfr/\mstar$), computed for the entire galaxy with the same timescale as \sigmasfr.

Although the distributions vary quite significantly between panels, at a given redshift, there is a substantial overlap in the range of values of both \sigmagasSF and \sigmasfr within the KS plane. The lowest and highest tails of these distributions at each redshift are typically dominated by galaxies within the lowest and highest stellar mass bins, respectively.
Overall, the distributions of galaxies of a given stellar mass quartile move within the KS parameter space towards lower values of \sigmagasSF and \sigmasfr with decreasing redshift: fewer and fewer galaxies are found far above the canonical KS relation\footnote{With the term canonical, we refer to the sequence of normal, star-forming discs, as defined in e.g. \cite{Daddi2010}, i.e. \sigmasfr $\propto$ $\Sigma_{\rm{gas}}^{1.4}$.} (solid line), at all stellar masses. This is accentuated after cosmic noon ($z<2$) where only a handful of galaxies reach the sequence of starbursts (dotted line).
%

As expected, the \ssfr of galaxies decreases with decreasing redshift, in particular at $z \leq 2$. It also decreases with increasing stellar mass at each redshift.
The \ssfr of galaxies strongly correlates with \sigmasfr at each redshift and in each stellar mass bin, essentially because the \ssfr is computed using stars with the same age as \sigmasfr ($< 10$ Myr). This strong correlation vanishes when considering older stars (e.g. $< 100$ Myr). As a consequence, at fixed \sigmagasSF, galaxies with shorter depletion times, have a higher \ssfr than those with longer depletion times. We stress that this behavior, although not surprising, is not obvious. Starbursting systems have short depletion times, i.e. the normalisation of the SFR by the gas mass\footnote{As such, lines of constant depletion time have a slope of unity in the KS plane. A slight difference exists with the observed sequence of starbursts which yields a slope of 1.4 in the KS plane \citep{Daddi2010}. This nuance is yet to be understood.}, while the \ssfr is the SFR normalized by the {\it stellar} mass. Galaxies with a given sSFR but different gas fractions could then have significantly different depletion times. As a matter of fact, the systematic qualification of starburst galaxies as outliers above the main sequence of star formation is being questioned by observations (\citealt{Gomez2022, Ciesla2023}, see also e.g. \citealt{Tacconi2018}) and simulations \citep{Renaud2022}.

The trends and correlations from Fig.~\ref{fig:KS_SFR10_gasSF_ssfr} persist when the neutral gas (\sigmagastot) is considered instead of the molecular gas alone (\sigmagasSF, see Fig.~\ref{fig:KS_SFR10_gastot_ssfr}), but with a steepening of the slopes, weakening of correlations, and increased scatter, in agreement with observations, at all redshifts and in all stellar mass bins. 

Figure~\ref{fig:KS_ssfr_gas_SF_tot} shows the distributions in the \sigmagasSF -- \sigmasfr and  \sigmagastot -- \sigmasfr planes, but now for the entire population of galaxies at different redshifts, by stacking all galaxies from Figs.~\ref{fig:KS_SFR10_gasSF_ssfr} and \ref{fig:KS_SFR10_gastot_ssfr}, respectively, where 2D histograms are computed by averaging the colour-coded quantity in each bin.
The gradients in \ssfr seen in individual stellar mass bins are still apparent when stacking all stellar masses.
Similarly, the entire galaxy population shows a stronger correlation, smaller dispersion, and shallower slope between the SFR density and the molecular gas, than with the neutral gas.

The shallower slope of the correlation with \hmol is consistent with studies of nearby galaxies both on galactic \citep[e.g.][]{Kennicutt1998,Liu2015,delosReyes2019,Kennicutt2021} and sub-kpc scales \citep[e.g.][]{Kennicutt2007,Bigiel2008,Leroy2008}. Although the value of the slope depends on the employed fitting method and types of galaxies under consideration, it is found to be approximately linear.
The relation between SFR and total gas surface densities for a combined sample of normal and starburst galaxies is found to be superlinear with slopes $1.4-1.5$ \citep[][]{Kennicutt1998,Kennicutt2021}. A similar slope is found for a sample of nearby normal spiral galaxies \citep[][]{delosReyes2019} and at higher redshifts \citep[$z\sim1.5$;][]{Daddi2010}, while the inclusion of dwarf galaxies tends to produce a shallower slope of $\sim 1.3$ \citep{delosReyes2019}.  
When fitted separately, starburst galaxies appear to follow a relation with slope $1-1.2$, as recently revealed by \cite{Kennicutt2021}, which is shallower compared to previous studies finding slopes of $1.3-1.4$ \citep[e.g.][]{Kennicutt1989,Daddi2010}, but confirms bimodal (or possibly multimodal) relation for the global star formation \citep[e.g.][]{Daddi2010,Genzel2010}.

We will now explore when these observed relations emerge.

\subsection{Emergence of the KS relation}
\label{subsec:emergence}

\begin{figure}
\centering\includegraphics{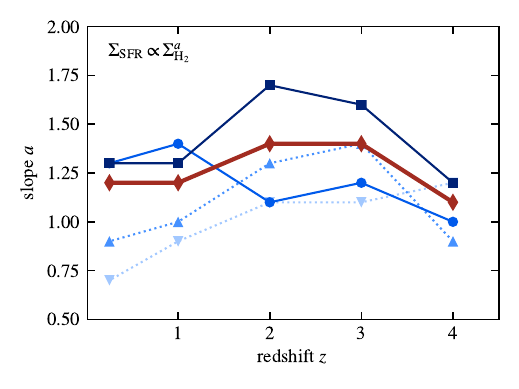}\\
\includegraphics{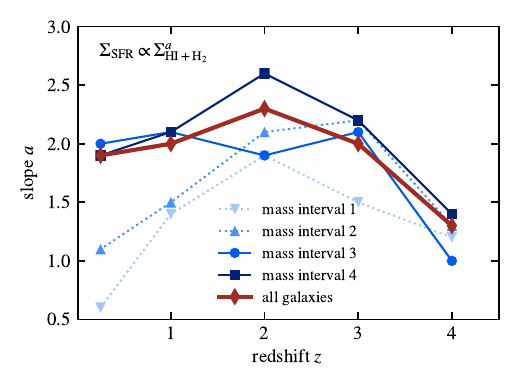}
\caption{Evolution of the slope of the fits of the KS relation from Figs.~\ref{fig:KS_SFR10_gasSF_ssfr} and \ref{fig:KS_SFR10_gastot_ssfr}, i.e. using the Bayesian fitting method, for the four mass bins considered, from the low mass bin in a light color to the most massive one in dark color. The red points show the slope of the entire galaxy population at a given redshift (i.e. without accounting for their mass, as in Fig.~\ref{fig:KS_ssfr_gas_SF_tot}). The emergence of the KS relation is shown by the convergence of the slope of the massive galaxies (from the two most massive bins) near the observed relation at $z\approx 2\--3$. Low mass galaxies do not show signs of convergence toward a fixed slope: their KS relation gets continuously shallower after $z\approx 2\--3$.}
\label{fig:emergence}
\end{figure}

\begin{figure}
\centering\includegraphics{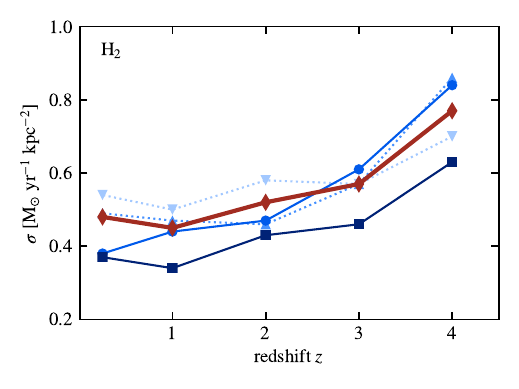}\\
\includegraphics{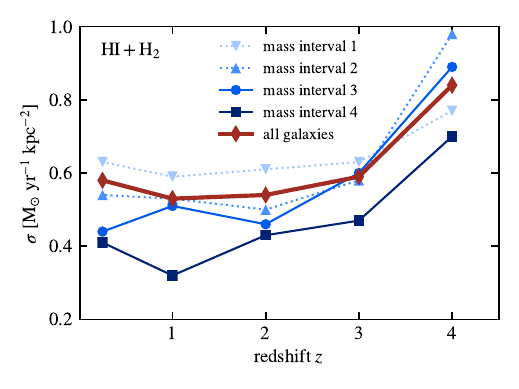}
\caption{Same as Fig.~\ref{fig:emergence}, but showing the dispersion around the best fit. Only the vertical dispersion, i.e. in $\log(\sigmasfr)$ is considered here.
}
\label{fig:emergence_disp}
\end{figure}

The redshift dependence of the trends highlighted in the previous section suggests that the KS relation evolves with cosmic time. In this section, we explore its emergence and overall evolution. Fig.~\ref{fig:emergence} shows the evolution of the slope $a$ of the relations ($\Sigma_{\rm SFR}\propto\Sigma_{\rm gas}^a$) fitted with the Bayesian linear regression method, and Fig.~\ref{fig:emergence_disp} displays the dispersion of the data around these fits (see Appendix~\ref{sec:emergence_altfit} for the equivalent plots using another fitting method).

The least massive galaxies (the lower half of the stellar mass distribution) clearly differ from the most massive cases (the highest stellar mass bin): at all redshifts, their KS relations are shallower and more dispersed. 
An examination of the distributions (Fig.~\ref{fig:KS_SFR10_gasSF_ssfr}) reveals that this originates from the presence of galaxies with short depletion times at low $\sigmagasSF$.
Such cases of rapid star formation in (relatively) diffuse gas possibly due to environmental triggers like mergers, or strong (or fast) gas outflows, are found at low mass at all redshifts, but only at high redshift for the massive quartiles. At cosmic noon ($z\sim 2$), this regime disappears at all masses but reappears at $z\lesssim 2$ at low masses. This explains the overall ``bell'' shape at low mass in Fig.~\ref{fig:emergence}. This effect is significantly more pronounced in the neutral gas (\hi + \hmol), which indicates that the {\it fraction} of molecular over neutral gas plays a role in star formation in diffuse gas. It is likely that at the lowest masses, the galaxies comprise only one active star-forming region at a given instant, i.e. a molecular cloud with an extended atomic envelope (recall Fig.~\ref{fig:mosaic}), which would favour the star formation regime noted here.

The slope of the KS relation stabilizes below $z\approx2$ for the overall population (red thick line on Fig.~\ref{fig:emergence}) and the most massive galaxies ($M_{\star} \gtrsim 10^{8} {\rm M}_{\odot}$ at this redshift). This is also the epoch when the dispersion around the relation reaches its final, minimum plateau (Fig.~\ref{fig:emergence_disp}). Therefore, the present-day KS relation emerges at cosmic noon ($z\approx 2$) in the most massive galaxies. Our results predict that populating the KS plane with observational data from the top 50\% most massive galaxies at redshifts $\gtrsim 3$ would result in a different and significantly more dispersed relation than the one currently established at low and intermediate redshifts (when considering local spirals, $z=1\-- 2$ discs, and BzK galaxies, \citealt{Daddi2010,Tacconi2010,Salmi2012}).

However, this is not the case for the least massive galaxies, for which no stabilization of the slope is seen, neither in molecular nor neutral gas. Interestingly, the dispersion around the best fit of these galaxies still yields a behaviour very similar, qualitatively and quantitatively, to that of the most massive ones, i.e. a decrease until $z\approx 2-3$ followed by a relatively flat plateau. Hence, the relation for the low mass galaxies becomes simultaneously tighter and shallower below $z\approx 2-3$. This could indicate that the extreme cases at high redshift either evolve to a more massive quartile via rapid growth or conversely get quenched and disappear from the star-forming sample. The remaining low-mass objects would then display a more homogeneous behavior.

These shallower KS relations of low-mass galaxies are in qualitative agreement with the observations of local dwarf galaxies, which report slopes around unity \citep[e.g.][]{Filho2016, Roychowdhury2017}\footnote{As mentioned in the previous section, a quantitative agreement cannot be reached due to the diversity of fitting methods employed in the literature.}. The underlying reason is still debated and probably consists of an interplay between galaxy interactions and the low-metallicity contents of these dwarf galaxies which caps their efficiency at forming molecular gas \citep{Cormier2014}. Such hypotheses are in line with our measurements of star formation in diffuse gas that we interpret as star-forming regions with extended atomic envelopes. Confirming these ideas requires a resolved analysis of these galaxies, instead of the statistical approach we follow here. Thus, we will explore these hypotheses in a forthcoming paper.

When considering the relation between \sigmasfr and the neutral gas surface density \sigmagastot, we retrieve qualitatively the same evolution of the slope and the dispersion with redshift and stellar mass (for all mass quartiles), but with larger slopes. The reason for this steepening of the relations is the presence of sub-efficient star-forming regions in galaxies at low \sigmagastot, often referred to as the ``break'' of the KS relation (see an illustration in \citealt{Bigiel2008}). The physical origin of the break has been shown analytically \citep{Renaud2012} and numerically \citep{Kraljic2014} to be caused by low levels of turbulence which do not efficiently promote the formation of dense gas, or in other words, by a low filling factor of star-forming gas in the volumes considered. In turn, the break becomes more apparent when including the atomic component in our analysis, by increasing the gas surface density without altering \sigmasfr, which bends the distribution of galaxies in the KS plane below the canonical KS relation.

At low redshifts, the low mass galaxies of our sample correspond to dwarfs of which the low-metallicities \citep{Dubois2021} could explain the inefficient formation of molecular gas (at small scales), and in turn slow down star formation, even at high \sigmagasSF at galactic scales (but see the discussion of \citealt{Roychowdhury2017} on the relatively small effect of the metallicity on the KS relation). Interestingly, Figure 13 of \citet{Dubois2021} indicates that the relation between the stellar mass and the metallicity in \newh varies only very weakly with the redshift. We have checked that this remains true when selecting the star forming galaxies only. We confirm that the gas and stellar metallicities\footnote{The metallicity is computed within the $R_{\rm eff}$. For the gas metallicity, only the neutral phase is considered.}
for the selection of galaxies within the KS plane increase with stellar mass and with decreasing redshift, as expected, but the redshift evolution of the mass-metallicity relation is only weak. Moreover, at all redshifts, at a given stellar mass, the metallicity does not show any gradient within the KS plane. This implies that low redshift dwarfs have a similar metallicity as the galaxies with the same stellar mass at $z \gtrsim 2$, but which are then in our upper mass bin, and already follow distributions close to the canonical KS relation. This demonstrates that the stellar mass and the metallicity are not key parameters in driving the emergence of the KS relation. The role of other physical quantities is explored in the next section.

\subsection{Physical drivers of the KS relation}

\subsubsection{Molecular and total gas content}
\label{subsec:gas}

\begin{figure*}
\centering\includegraphics[width=0.9\textwidth]{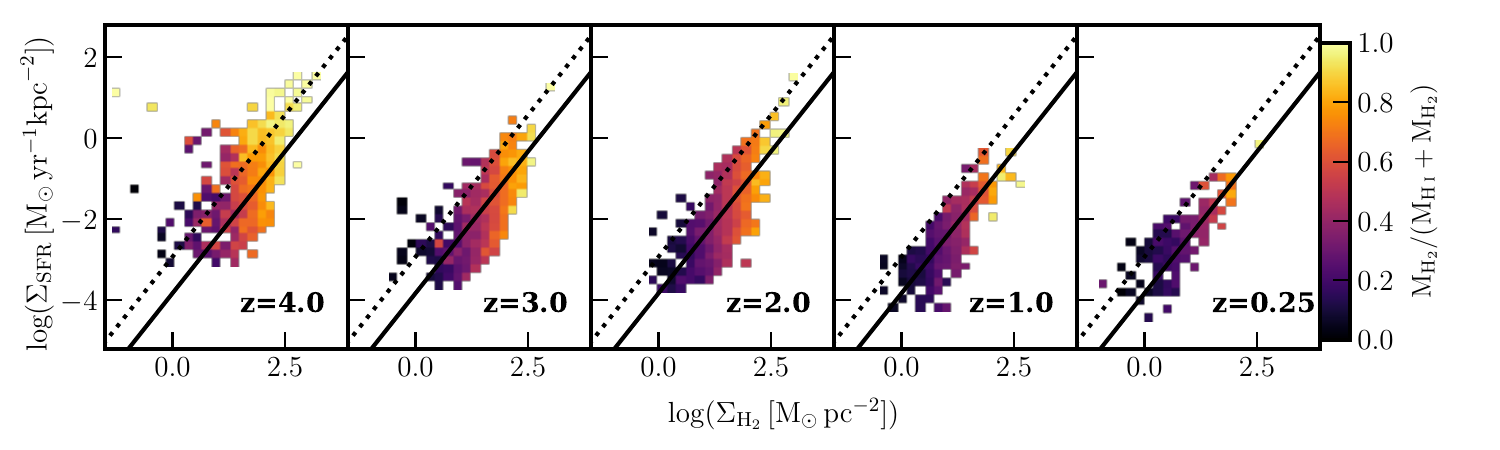}
\centering\includegraphics[width=0.9\textwidth]{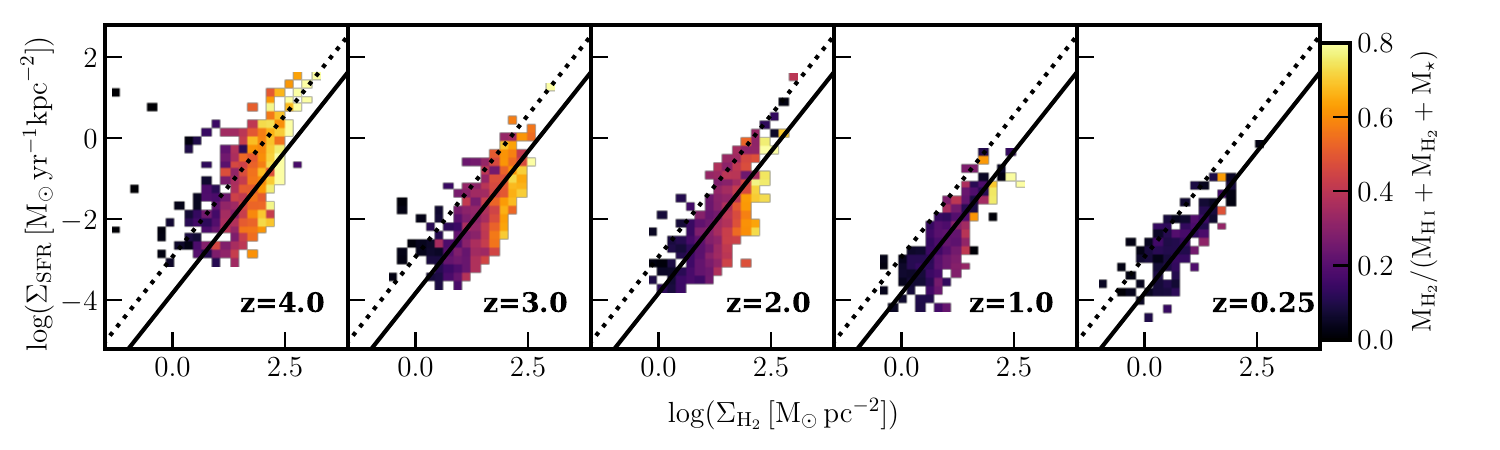}
\centering\includegraphics[width=0.9\textwidth]{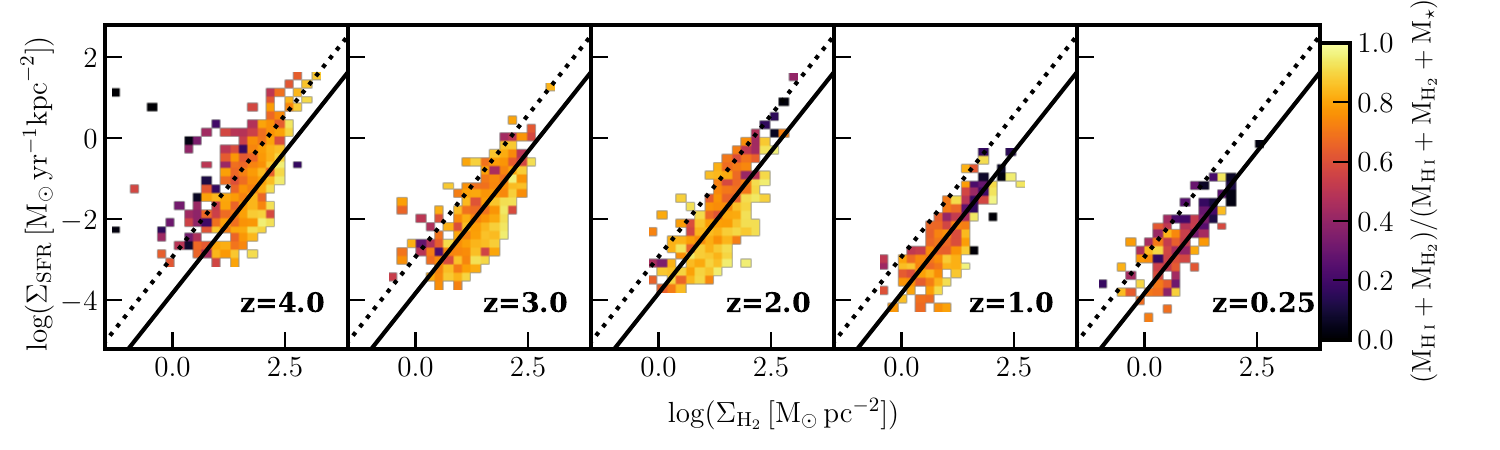}
\\[-0.1cm]
\caption{Same as Fig.~\ref{fig:KS_ssfr_gas_SF_tot}, but color-coded by the molecular over cold gas fraction (top), and baryonic molecular (middle) and neutral (bottom) gas fractions. Molecular gas fraction (top panel) correlates with \sigmagasSF at all redshifts. Baryonic molecular fraction (middle panel) correlates with \sigmagasSF at high redshifts, below z$\sim$2 this correlation weakens and is essentially carried by low-mass galaxies. The correlation between the neutral gas fraction and \sigmagasSF is apparent only at $z=4$. At $z\lesssim$2 the trend reverses such that this fraction decreases with increasing \sigmagasSF.}
\label{fig:KS_fgas_SF_gas_both}
\end{figure*}

We now investigate whether the relation between \sigmagasSF and \sigmasfr is driven by the gas content of galaxies.

Figure~\ref{fig:KS_fgas_SF_gas_both} (top panel) shows the molecular gas fraction, defined as the fraction of \hmol mass over the neutral gas, i.e. M$_{\rm{H}_2}$/(M$_{\textsc{H\,i}}+\rm{M}_{\rm{H}_2}$), and its evolution within the KS plane with the cosmic time for the entire galaxy population. The fraction of molecular gas decreases with decreasing redshift. Moreover, it correlates with \sigmagasSF resulting in vertical contours (within the KS plane) at all redshifts. 
The same trends are seen at a given stellar mass, although the fraction of molecular gas increases with stellar mass (Fig.~\ref{fig:KS_SFR10_gasSF_fgas_SF_overtot}, see also \citealt{Dubois2021}, their figure 19). 

Molecular gas content of galaxies may also be defined in terms of baryonic fraction, i.e.  M$_{\rm{H}_2}$/($\rm{M}_{\textsc{H\,i}} + \rm{M}_{\rm{H}_2} + \rm{M}_\star$). The correlation between the baryonic molecular gas fraction and \sigmagasSF is maintained at all redshifts (Fig.~\ref{fig:KS_fgas_SF_gas_both}, middle panel), although it weakens at $z \lesssim 2$ when it is only carried by the low-mass galaxies -- the massive galaxies having very low baryonic molecular gas fraction, independently of \sigmagasSF (see Fig.~\ref{fig:KS_SFR10_gasSF_fgasSF}).
This lack of correlation at low redshift results from the baryonic fraction of molecular gas vs stellar mass relation getting shallower at these late times \citep[see top right panel of figure 19 of][]{Dubois2021}.

We finally consider the neutral baryonic gas fraction, i.e. (M$_{\rm \textsc{H\,i}} + \rm{M}_{\rm{H}_2})/(\rm{M}_{\textsc{H\,i}} + \rm{M}_{\rm{H}_2} + \rm{M}_\star)$. As already reported by \citet{Dubois2021} for the \newh galaxies, this fraction strongly anticorrelates with the stellar mass, but only mildly depends on the redshift at a given mass. This is confirmed in the KS plane for galaxy stacks (Fig.~\ref{fig:KS_fgas_SF_gas_both}, bottom panel) and individual stellar mass bins (Fig.~\ref{fig:KS_SFR10_gasSF_fgastot}). 
At high redshift ($z=4$), the neutral gas fraction increases with \sigmagasSF for all stellar masses, but this correlation disappears at later epochs, where this fraction varies weakly with \sigmagasSF at fixed stellar mass but varies strongly with stellar mass. The combination of these relations between the neutral gas fraction and \sigmagasSF at high $z$, and \mstar at all $z$, translates into a non-trivial evolution of the distributions of the neutral gas fraction in the KS plane when the stellar mass is marginalized out. The reversal of the trend between the neutral gas fraction and \sigmagasSF between high and low redshift is thus a direct consequence of the dependencies highlighted above, and of how early the galaxies build up their stellar masses.

In conclusion, both the molecular and neutral gas fractions vary with one of the parameters of the KS plane (\sigmagasSF), but have close to no influence on the other (\sigmasfr), except in the very diffuse gas of the low mass galaxies, as noted above. As such, at galactic scales, the KS relation does not originate from the gas fractions of the galaxies, at any redshift, nor at any stellar mass. 
The quantity that correlates better within the KS plane is the fraction of cold gas that is in the dense phase, however, it does not fully capture the variation with both the surface density of gas and the star formation rate of galaxies.

\subsubsection{Turbulence}
\label{subsec:Mach}

\begin{figure*}
\centering\includegraphics[width=0.9\textwidth]{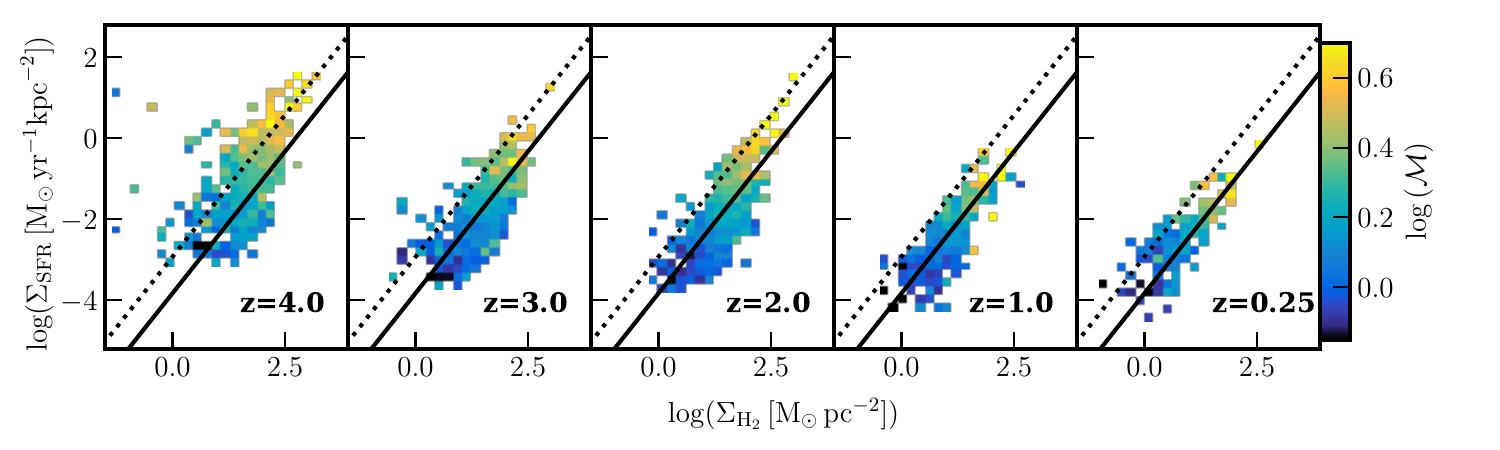}
\caption{Same as Fig.~\ref{fig:KS_ssfr_gas_SF_tot}, but colour-coded by the Mach number \mach. At all redshifts, \mach increases monotonically with increasing both \sigmagasSF and \sigmasfr. At fixed \sigmagasSF, \mach is correlated with \sigmasfr.
}
\label{fig:KS_Mach_gas_SF}
\end{figure*}

\begin{figure*}
\centering\includegraphics[width=0.33\textwidth]{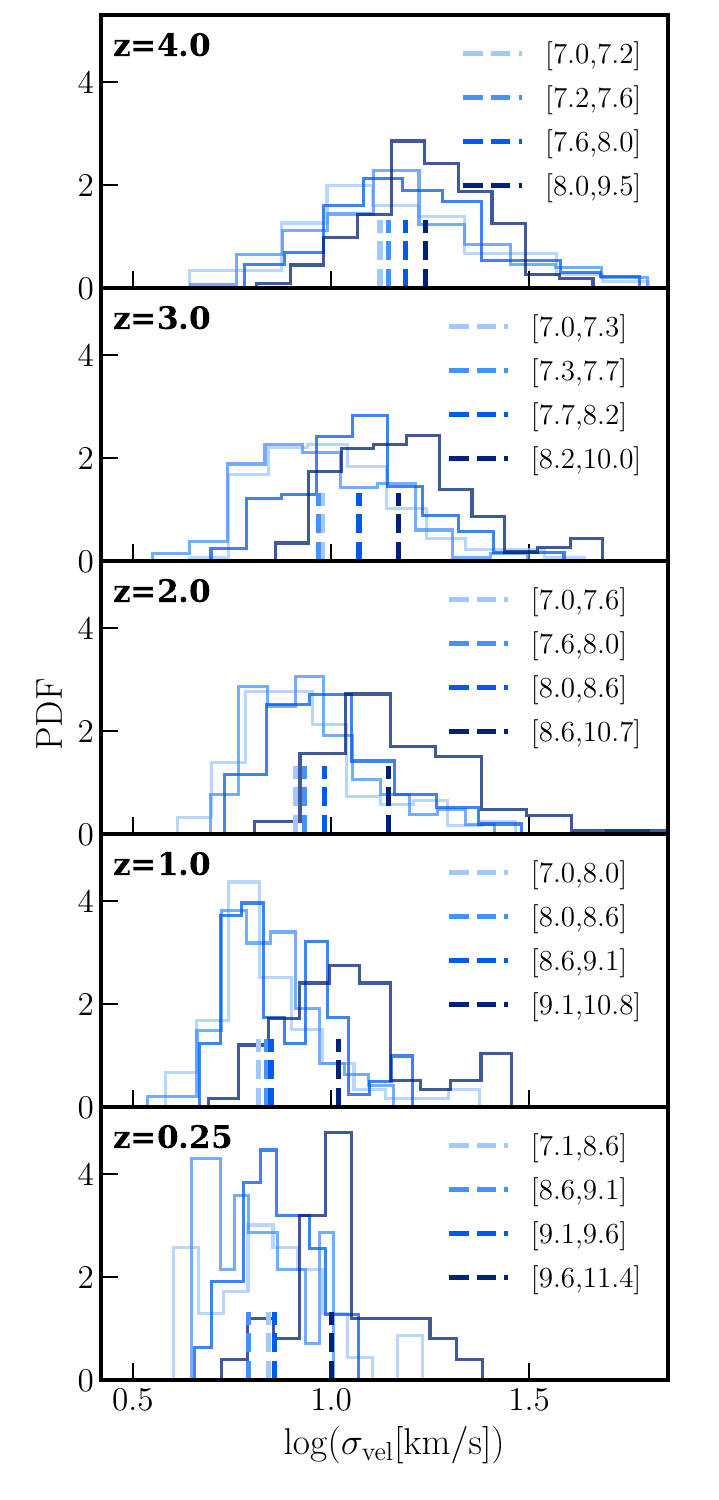}
\centering\includegraphics[width=0.33\textwidth]{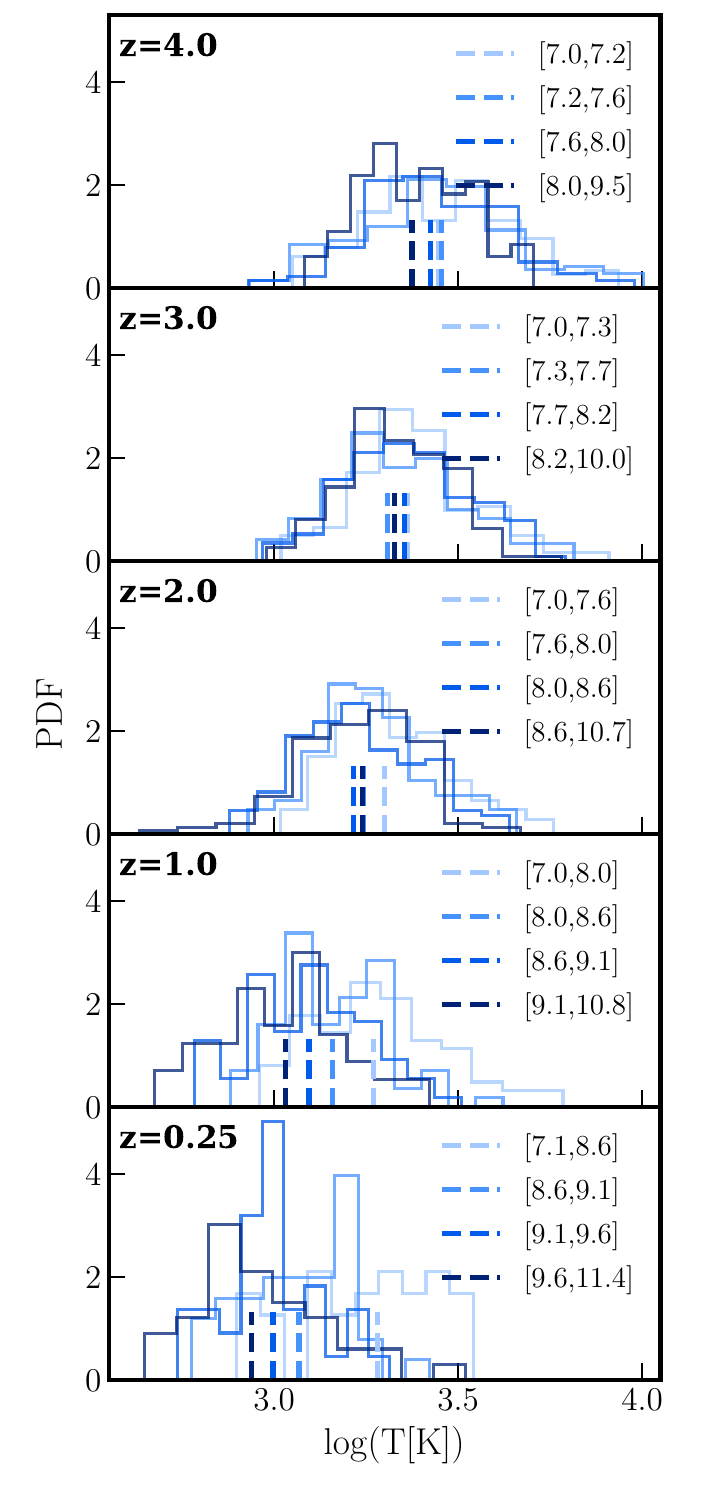}
\centering\includegraphics[width=0.33\textwidth]{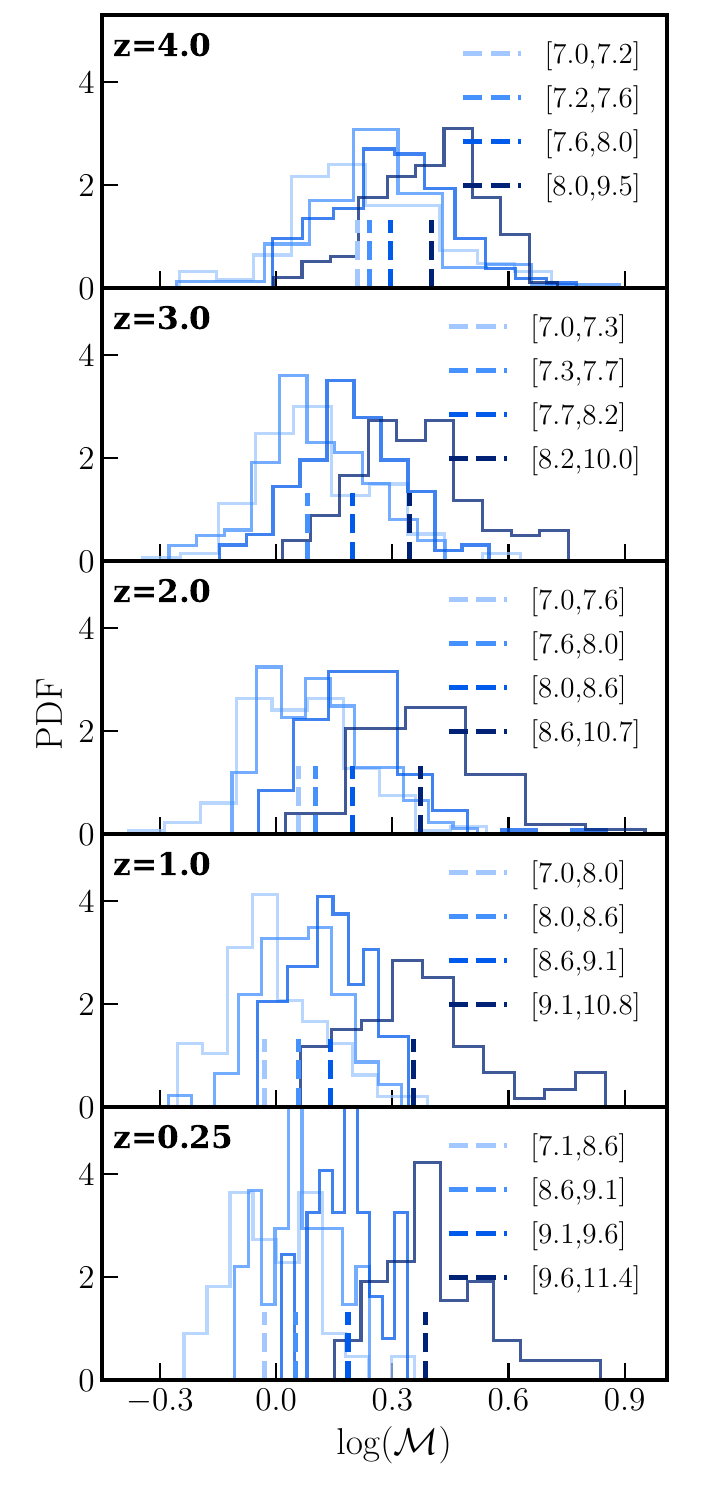}
\caption{Normalised distributions of gas velocity dispersion (left), temperature (middle), and Mach number (right) in different stellar mass bins (colored lines, as indicated in the legend, reporting the logarithm of \mstar in units of \msun) and redshifts (indicated in the left upper part of each panel). Vertical lines represent medians of distributions. 
}
\label{fig:hist}
\end{figure*}

Previous works \citep[e.g.][]{Vazquez1994,Renaud2014} have pointed out the role of turbulence in setting the density distributions of gas, and thus the amount of star-forming gas in galaxies. Here, we test whether turbulence explains the emergence of the KS relation over cosmic time.

Figure~\ref{fig:KS_Mach_gas_SF} shows the distribution of galaxies in the KS plane, now color-coded with their turbulence Mach number \mach. At all redshifts, \mach evolves monotonically along the best-fit relation in the KS parameter space, by increasing with \sigmagasSF and \sigmasfr. Furthermore, at fixed \sigmagasSF, \sigmasfr is positively correlated with \mach.

As shown in Fig.~\ref{fig:KS_SFR10_gastot_Mach}, these trends are independent of the galaxy's stellar mass. At fixed stellar mass, galaxies at high redshifts are more turbulent than their low redshift counterparts, and at fixed redshift, more massive galaxies tend to be more turbulent than their lower mass counterparts, a direct consequence of increasing gas richness in galaxies with redshift \citep[e.g.][]{Bournaud2010,Renaud2012}.

To investigate the physical origin of these trends, Fig.~\ref{fig:hist} shows the distributions of the Mach number $\mach \propto \sigma_{\rm vel} / \sqrt{T}$, and the underlying quantities which are the gas velocity dispersion ($\sigma_{\rm vel}$) and the temperature ($T$)\footnote{As in the case of velocity dispersion and Mach number, the temperature is also computed as a mass-weighted average of gas cells within the $R_{\rm eff}$. Therefore, the temperature values are not directly comparable to typical values within individual molecular clouds.}. 

While the velocity dispersion decreases with redshift for all mass bins, only massive galaxies maintain high values at low $z$ ($\sim 10\ \rm{km\ s}^{-1}$), leading to significantly higher median values. This is confirmed by observations at low redshifts of lower dispersion in dwarf galaxies  ($\sim 1\ \rm{km\ s}^{-1}$) than in massive galaxies ($\sim 10\ \rm{km\ s}^{-1}$, e.g. \citealt{Hunter2021}). This general trend likely originates in parts from the overall lowering of the star formation activity with cosmic time, and possibly the less efficient coupling of feedback with the local interstellar medium (ISM), as opposed to intergalactic medium due to low escape velocity, in low-mass galaxies (i.e. with shallow potential wells).

The locations of the galaxies with high $\sigma_{\rm vel}$ in the KS plane (Fig.~\ref{fig:KS_SFR10_gasSF_vsig}) reveal complex correlations with the indicators of star formation: while galaxies with short depletion times tend to have high $\sigma_{\rm vel}$, high-velocity dispersion are found across the entire KS plane. This indicates that stellar feedback is not the only factor in setting the velocity dispersion, and therefore the KS relation~\citep[see also][]{Agertz2011,Agertz2015}, at the galactic scale.
Galactic dynamics and interaction-triggered stirring are likely important drivers of the velocity dispersion (see \citealt{Renaud2014} for an illustration that increased velocity dispersion is a cause and not a consequence of starbursts in mergers).

The temperature of all mass quartiles decreases with decreasing redshift, but Fig.~\ref{fig:hist} (middle-column) reveals that the stellar mass only discriminates the distributions of $T$ at late times ($z \lesssim 1$). Contrary to the velocity dispersion, there is no relation between the star formation indicators and the temperature in the KS plane (Fig.~ \ref{fig:KS_SFR10_gasSF_T}), which is consistent with the interpretation of the limited impact of feedback, even though the details, in particular on small scales, might be more complicated.

In terms of Mach number, the trends noted from the two underlying quantities ($\sigma_{\rm vel}$ and $T$) naturally combine to lead to a shift of the distributions of $\mach$ towards low values with decreasing redshift, an effect which is significantly more pronounced for low-mass galaxies (Fig.~\ref{fig:hist}). 

A more detailed examination (Fig.~\ref{fig:KS_SFR10_gastot_Mach}) reveals that the trends found in velocity dispersion and in temperature conspire to give rise to a clear evolution of $\mach$ along the KS relation, with `tighter correlation' with $\Sigma_{\rm gas}$ and $\Sigma_{\rm SFR}$
than the individual $\sigma_{\rm vel}$ and $T$. This further demonstrates the paramount role of turbulence in the star formation activity, in particular in the KS plane.

Higher Mach numbers favor higher density contrasts in the ISM (i.e. a wider gas density PDF, see \citealt{federrath2008}), and thus the formation of a larger faction of dense molecular gas. This explains that dwarf galaxies at low redshift, with a low Mach number, tend to have lower molecular gas fractions (Fig.~\ref{fig:KS_SFR10_gasSF_fgas_SF_overtot}) than their massive counterparts, and thus appear below the canonical KS relation, in the so-called ``break'' (see \citealt{Kraljic2014}), when considering  the total neutral gas (Fig.~\ref{fig:KS_SFR10_gastot_ssfr}), but are shifted toward the low gas densities when considering the molecular phase only (Fig.~\ref{fig:KS_SFR10_gasSF_ssfr}). Finally, the high \sigmasfr of these galaxies implies that this shift toward low \sigmagasSF places them above the canonical KS relation, which drives the flattening of the relation of these sub-populations (recall Fig.~\ref{fig:emergence}). 

Furthermore, the histograms of Fig.~\ref{fig:hist} show that the distributions of velocity dispersion in low-mass galaxies become peaked toward the low-value end at low redshift, while the massive galaxies only exhibit a tail with only a few cases at such low-velocity dispersion, and the bulk of their distribution remains centered around higher values ($\sim 10\ \rm{km\ s}^{-1}$) with little evolution after $z \lesssim 2$. In other words, the lower end of the distributions in $\sigma_{\rm vel}$ gets more and more populated with low-mass galaxies with decreasing redshift, while the distribution of velocities dispersion of massive galaxies ceases to evolve (statistically). For the reasons discussed before, this transpires in the histograms of Mach number, and finally in the distribution of galaxies in the KS plane. Therefore, the mass-dependent evolution of the velocity dispersion explains the convergence of the slope of the KS relation at high mass after $z \approx 2$, and the absence of the convergence in low-mass galaxies, noted in Fig.~\ref{fig:emergence}.

\section{Discussion}
\label{sec:discuss}

\subsection{Scale and projection effects}

\begin{figure*}
\centering\includegraphics[width=0.98\textwidth]{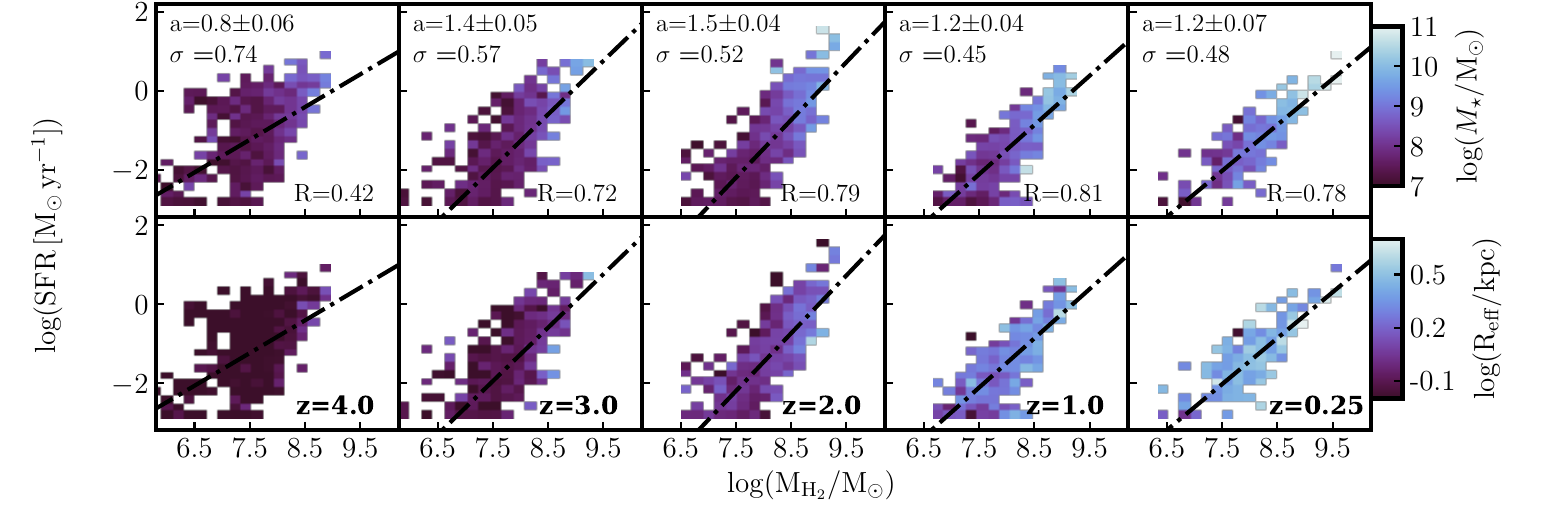}
\caption{Distribution of galaxies in the \mgassf-\sfr plane, i.e. deprojected version of the KS plane, at redshifts 4, 3, 2, 1 and 0.25, from left to right, respectively,  as a function of galaxy mass (top) and effective radius (bottom).
Dash-dotted lines are fits in the logarithmic space at each redshift, with the slope $a$ and the standard deviation of residuals $\sigma$. The coefficient $R$ is the Pearson correlation coefficient.
At all redshifts, more massive galaxies tend to have higher \sfr and \mgassf compared to their lower mass counterparts. As galaxies grow in mass, they grow in size, however, no obvious correlation is seen between \effr, \sfr and molecular gas mass.
}
\label{fig:KS_int_SFR10_gasSF_reff}
\end{figure*}

So far, we have conducted our analysis using the observables \sigmagasSF, or \sigmagastot, and \sigmasfr, and identified relations in the KS plane. However, by doing so, we effectively introduce an arbitrary choice for the spatial scale used in the measurement of both quantities, which necessarily impacts the values of the surface densities and possibly artificially distorts the distributions of galaxies in the KS plane. To establish whether our conclusions depend on our choice of examining projected quantities, and over the scale of the effective radius, we plot in Fig.~\ref{fig:KS_int_SFR10_gasSF_reff} the distributions of galaxies of our sample in the plane of molecular gas mass vs. \sfr, i.e. a deprojected version of the KS plot, colour-coded by mass (top) and size \effr (bottom). At all redshifts, more massive galaxies tend to have higher \sfr and molecular gas mass, however, there is no obvious correlation between \sfr, \mgassf, and effective radius of galaxies. Furthermore,  all the trends (or the absence thereof) with molecular and total gas fractions, and \mach seen in the KS parameter space are retrieved with deprojected quantities, as shown in Fig.~\ref{fig:KS_int_SFR10_gasSF_3x5}. Therefore, the trends seen in the KS plane are not primarily driven by measuring the physical quantities in projection rather than in 3D, and the scale adopted does not introduce biases in the distributions.

The diversity of star formation activities seen in the wide distribution of \sigmagasSF, \sigmagastot, and \sigmasfr, but also in the slopes and the scatters of the KS relation, results from the convolution of two other effects: (i) the diversity of galaxies in the sample, illustrated by the variations of the KS relations with  stellar mass and redshift (Fig.~\ref{fig:KS_SFR10_gasSF_ssfr}), and (ii) the integration of the local, small-scale star formation law over entire galactic scales where not all the ISM is star-forming. Indeed, the shape (slope, offset, break, scatter) of the distributions of galaxies in the KS plane is driven by the underlying distribution of the physical properties of the star-forming regions within each galaxy, and it is the evolution of these distributions as functions of redshift, galactic mass, and  other factors like the environment, that sets the evolution of the KS relations. 

\subsection{Sub-grid models}

Our work points out the key role of turbulence (Mach number) in driving the KS relation at the galactic scale, already at high redshift. This confirms the analytical results of \cite{Renaud2012}, and the numerical work of \cite{Kraljic2014} which conducted a similar study without cosmological context, and for galaxies in the nearby Universe only (i.e. at low gas fractions). 
We note that the star formation model in \citet{Kraljic2014} differs from those in \newh, as it uses a fixed star formation efficiency per free-fall time and is applied at a higher resolution ($\sim$ 1 pc). In other words, the right-hand side of Equation~\ref{eq:sfr_ff} reduces to a constant value in the former. The paramount role of turbulence in setting the KS relation  found with both models (independently of an explicit inclusion of turbulence in the star formation model in the latter), further strengthens our conclusion.

Yet, it is important to keep in mind that the differences in the sub-grid models used do not necessarily correspond to different physics. Schemes that do not capture the formation of star-forming clouds (i.e. at resolutions $\gtrsim 20 $ pc) ought to incorporate this aspect in their star formation prescription. This can be achieved by imposing a criterion on the instability of the gas, through e.g. converging flows and/or the virial parameter, as is done in \newh. However, at cloud scale ($\approx 1\-- 10$ pc), the fragmentation of the ISM into individual clouds is already captured by the simulations, such that an instability criterion is redundant: at high resolution, the dense gas has necessarily gone through the instability phase. As for any physical mechanism, it is crucial to identify which processes are not captured explicitly by the simulation, in order to construct and parametrize the ``sub'' aspect of the sub-grid models.

By conducting a spatially resolved study of the KS relation within the {\sc Fire} \citep{Hopkins2014} framework, \cite{Orr2018} pinpointed instead the central role of stellar feedback in regulating star formation at small scales \citep[see also][for similar conclusion for the global KS relation]{Dekel2019}. We note however that their sub-grid treatment de facto implies an important role for feedback, as it is required to regulate the star formation process set with an efficiency of 100 percent (as opposed to $\sim 0.1-10 \%$ in most of the literature from observations and simulations). It is, therefore, possible that a different conclusion, perhaps closer to ours, would be reached by adopting a lower star formation efficiency, and thus automatically decreasing the impact of feedback \citep[see e.g.][]{Brucy2020,Brucy2023,Hu2023}. For instance, in the analytical model of \cite{Renaud2012}, stellar feedback can be introduced as a cap on the effective star formation efficiency. While this results in  lowering  the slope of the KS relation, a requirement to match observations, a KS-like power-law does exist without feedback, and originates solely from the log-normal shape of the distribution of gas density, itself known to be set by turbulence \citep[e.g.][]{Vazquez1994}. The fact that the KS relation can be modeled from such different underlying physics suggests that divergences should be sought in more fundamental quantities or behaviour, and possibly at smaller scales, i.e. before the differences between models get blended in integrated, projected, and galactic-scale averaged measurements. Considering other, more extreme environments where the relative contributions of the mechanisms involved vary could certainly provide interesting insights.

\subsection{A variety of possible underlying physical mechanisms}

Taking advantage of the broad diversity of galaxies in \newh, we show here that the gas fraction does not strongly influence the star formation relation, at least when integrated over entire galaxies, and that the mild trends found between the gas fraction and its surface density (be it neutral or molecular) actually get reversed at $z \approx 2$. This change entails an evolution of the morphology and size of the star-forming volume, likely connected to a more concentrated activity. Underlying physical reasons could be the intrinsic evolution of discs towards fueling more and more gas to the nuclei, and/or environmental effects due to gravitational torques exerted on disc material by more and more massive companion galaxies.

Some of these aspects have been explored in the special case of the Milky Way-like galaxy, using the \vintergatan simulation \citep{Agertz2021, Renaud2021}. These works highlight the necessity for the galactic disk to be in place \citep{Park2021} for the galaxy to strongly react  via large-scale  wakes to interactions through a starburst activity \citep{Segovia2022}. The redshift-dependence of tidal compression, both in terms of intensity and mass involved also appears as crucial in the cosmic evolution of starbursts \citep{Renaud2022} as it controls energy input at the cascade injection scale. Exploring these points further and over an entire population of galaxies, like that in \newh, requires a dedicated analysis of the diversity of individual evolution that builds the population statistics shown here, which we leave for a forthcoming paper (Kraljic et al. in preparation).

Galaxies with the highest {\it global} turbulence level are not only those which host the densest gas and form the most stars (as shown by their locations at high surface densities of gas and of SFR). They are also the galaxies with the shortest depletion time. This is particularly visible at high redshift ($z \gtrsim 2$) in Fig.~\ref{fig:KS_Mach_gas_SF}, where the most turbulent galaxies lie around the starburst relation from \cite{Daddi2010}, about 1 dex above the canonical KS relation. This is in line with the conclusions of \cite{Renaud2014,Renaud2019b,Renaud2021b} which reported that the increase of the level of compressive turbulence in mergers can lead to a starburst activity. In this context, it is crucial to differentiate the production of many stars (high SFR) from the fast production of stars (short depletion time). While the two aspects are independent, our results show that high levels of turbulence make some high redshift galaxies reach both a high \sigmasfr, and a short depletion time. This then changes at lower redshift, when the most turbulent galaxies of our sample do not necessarily yield the shortest depletion times. Such an evolution could be connected with the rarefaction of the major mergers at late epochs, and thus the statistical lowering of the tidal and turbulent compression \citep{Renaud2022}, and likely has implications on the evolution of the normalization of the main sequence of galaxy formation \citep[e.g.][]{Tacconi2018}.

\section{Conclusion}
\label{sec:conclusion}

We have investigated the distributions of galaxies from the cosmological hydrodynamic simulation \newh in the KS parameter space, as a function of their stellar mass and redshift, the emergence of the star formation scaling-laws at the galactic scale, and its physical drivers. Our main results are:

\begin{itemize}
\item Both the stellar mass and redshift influence the overall location of the galaxy population in the KS plane.
\item A power-law relation of the form $\sigmasfr \propto \Sigma_{\rm gas}^{a}$ with a slope $a\approx 1.4$ emerges at $z\approx 2\-- 3$ for the most massive half of the galaxy population ($\mstar \gtrsim 10^8\, \msun$ at these redshifts) in agreement with observations up to these redshifts. However, the slope of the relation varies at earlier epochs, with an increased scatter. This indicates that the KS relation might not provide a robust calibration for star formation in galaxies at very high redshift. For the least massive galaxies, there is no sign of the convergence of the slope of their distribution in the KS plane, as it continues to get shallower at the last epochs. The slopes are systematically higher when considering the total neutral gas as opposed to the molecular gas.
At all stellar masses, the dispersion around the best-fit relation decreases with the decreasing redshift.
\item The gas fraction (neutral or molecular) does not correlate with the star formation activity as traced by \sigmasfr and therefore does not play a primary role in establishing the KS relation.
Similarly, neither the velocity dispersion of the gas nor its temperature alone can fully explain the star formation activity of galaxies as captured by the KS relation, pointing towards a limited impact of feedback.
\item Conversely, the level of turbulence of the interstellar medium, as quantified by the Mach number, is found to drive the relation between gas and \sfr densities at all redshifts, independently of stellar mass. 
More specifically, it is the ability of a galaxy to reach a supersonically turbulent regime that matters, with the Mach number ($\mathcal{M} > 1$) being the driver of the KS relation independent of stellar mass.
At high redshift, for a given gas density, the most turbulent galaxies yield short depletion times, characteristic of starburst galaxies. Their frequency decreases at low redshift.
\end{itemize}

The evolution reported here and a number of previous works on the star formation activity at galactic scales point toward an important role of inflow, interactions, mergers, and the proximity of the disk to marginal stability in driving the star formation relations and their scatters. 
The latter could act as a confounding factor for efficient turbulent cascade and star formation, explaining  the emergence of tighter KS scaling relations, when secular dissipative processes take over.
Exploring these aspects requires tracking individual galaxies along their merger histories, and seeking changes in the properties of the star-forming material during the starburst phases, both at cloud and galactic scales. We will cover these topics in the forthcoming papers of this series.

\begin{acknowledgements}
This work was granted access to the high-performance computing resources of CINES under the allocations c2016047637 and A0020407637 from GENCI, and KISTI (KSC-2017-G2-0003). Large data transfer was supported by KREONET, which is managed and operated by KISTI. This work relied on the HPC resources of the Horizon Cluster hosted by Institut d’Astrophysique de Paris. We warmly thank S. Rouberol for running the cluster on which the simulation was post-processed. FR, OA, EA and ASO acknowledge support from the Knut and Alice Wallenberg Foundation and the Swedish Research Council (grant 2019-04659). FR acknowledges the support provided by the University of Strasbourg Institute for Advanced Study (USIAS), within the French national programme Investment for the Future (Excellence Initiative) IdEx-Unistra. EA acknowledges support from US NSF grant AST18-1546. SK acknowledges support from the STFC [grant numbers ST/S00615X/1 and ST/X001318/1] and a Senior Research Fellowship from Worcester College Oxford. TK was supported by the National Research Foundation of Korea (NRF) grant funded by the Korea government (No. 2020R1C1C1007079). S.K.Y. acknowledges support from the Korean National Research Foundation (2020R1A2C3003769). This work was supported in part by the Korean National Research Foundation (2022R1A6A1A03053472). This work is partially supported by the grant Segal ANR-19-CE31-0017 of the French Agence Nationale de la Recherche and by the National Science Foundation under Grant No. NSF PHY-1748958.
\end{acknowledgements}

%
%
\bibliographystyle{aa}
\bibliography{author.bib}

\begin{thebibliography}{98}
\expandafter\ifx\csname natexlab\endcsname\relax\def\natexlab#1{#1}\fi

\bibitem[{{Agertz} \& {Kravtsov}(2015)}]{Agertz2015}
{Agertz}, O. \& {Kravtsov}, A.~V. 2015, \apj, 804, 18

\bibitem[{{Agertz} {et~al.}(2021){Agertz}, {Renaud}, {Feltzing}, {Read},
  {Ryde}, {Andersson}, {Rey}, {Bensby}, \& {Feuillet}}]{Agertz2021}
{Agertz}, O., {Renaud}, F., {Feltzing}, S., {et~al.} 2021, \mnras, 503, 5826

\bibitem[{{Agertz} {et~al.}(2011){Agertz}, {Teyssier}, \& {Moore}}]{Agertz2011}
{Agertz}, O., {Teyssier}, R., \& {Moore}, B. 2011, \mnras, 410, 1391

\bibitem[{{Alves} {et~al.}(2007){Alves}, {Lombardi}, \& {Lada}}]{alvesetal07}
{Alves}, J., {Lombardi}, M., \& {Lada}, C.~J. 2007, \aap, 462, L17

\bibitem[{{Andersson} {et~al.}(2021){Andersson}, {Renaud}, \&
  {Agertz}}]{Andersson2021}
{Andersson}, E.~P., {Renaud}, F., \& {Agertz}, O. 2021, \mnras, 502, L29

\bibitem[{{Andr{\'e}} {et~al.}(2010){Andr{\'e}}, {Men'shchikov}, {Bontemps},
  {K{\"o}nyves}, {Motte}, {Schneider}, {Didelon}, {Minier}, {Saraceno},
  {Ward-Thompson}, {di Francesco}, {White}, {Molinari}, {Testi}, {Abergel},
  {Griffin}, {Henning}, {Royer}, {Mer{\'\i}n}, {Vavrek}, {Attard},
  {Arzoumanian}, {Wilson}, {Ade}, {Aussel}, {Baluteau}, {Benedettini},
  {Bernard}, {Blommaert}, {Cambr{\'e}sy}, {Cox}, {di Giorgio}, {Hargrave},
  {Hennemann}, {Huang}, {Kirk}, {Krause}, {Launhardt}, {Leeks}, {Le Pennec},
  {Li}, {Martin}, {Maury}, {Olofsson}, {Omont}, {Peretto}, {Pezzuto}, {Prusti},
  {Roussel}, {Russeil}, {Sauvage}, {Sibthorpe}, {Sicilia-Aguilar}, {Spinoglio},
  {Waelkens}, {Woodcraft}, \& {Zavagno}}]{andreetal10}
{Andr{\'e}}, P., {Men'shchikov}, A., {Bontemps}, S., {et~al.} 2010, \aap, 518,
  L102

\bibitem[{{Appleby} {et~al.}(2020){Appleby}, {Dav{\'e}}, {Kraljic},
  {Angl{\'e}s-Alc{\'a}zar}, \& {Narayanan}}]{Appleby2020}
{Appleby}, S., {Dav{\'e}}, R., {Kraljic}, K., {Angl{\'e}s-Alc{\'a}zar}, D., \&
  {Narayanan}, D. 2020, \mnras, 494, 6053

\bibitem[{{Aubert} {et~al.}(2004){Aubert}, {Pichon}, \& {Colombi}}]{aubert04}
{Aubert}, D., {Pichon}, C., \& {Colombi}, S. 2004, \mnras, 352, 376

\bibitem[{{Bigiel} {et~al.}(2008){Bigiel}, {Leroy}, {Walter}, {Brinks}, {de
  Blok}, {Madore}, \& {Thornley}}]{Bigiel2008}
{Bigiel}, F., {Leroy}, A., {Walter}, F., {et~al.} 2008, \aj, 136, 2846

\bibitem[{{Bigiel} {et~al.}(2011){Bigiel}, {Leroy}, {Walter}, {Brinks}, {de
  Blok}, {Kramer}, {Rix}, {Schruba}, {Schuster}, {Usero}, \&
  {Wiesemeyer}}]{Bigiel2011}
{Bigiel}, F., {Leroy}, A.~K., {Walter}, F., {et~al.} 2011, \apjl, 730, L13

\bibitem[{{Bolatto} {et~al.}(2013){Bolatto}, {Wolfire}, \&
  {Leroy}}]{Bolatto2013}
{Bolatto}, A.~D., {Wolfire}, M., \& {Leroy}, A.~K. 2013, \araa, 51, 207

\bibitem[{{Bouch{\'e}} {et~al.}(2007){Bouch{\'e}}, {Cresci}, {Davies},
  {Eisenhauer}, {F{\"o}rster Schreiber}, {Genzel}, {Gillessen}, {Lehnert},
  {Lutz}, {Nesvadba}, {Shapiro}, {Sternberg}, {Tacconi}, {Verma}, {Cimatti},
  {Daddi}, {Renzini}, {Erb}, {Shapley}, \& {Steidel}}]{Bouche2007}
{Bouch{\'e}}, N., {Cresci}, G., {Davies}, R., {et~al.} 2007, \apj, 671, 303

\bibitem[{{Bournaud} {et~al.}(2010){Bournaud}, {Elmegreen}, {Teyssier},
  {Block}, \& {Puerari}}]{Bournaud2010}
{Bournaud}, F., {Elmegreen}, B.~G., {Teyssier}, R., {Block}, D.~L., \&
  {Puerari}, I. 2010, \mnras, 409, 1088

\bibitem[{{Brucy} {et~al.}(2020){Brucy}, {Hennebelle}, {Bournaud}, \&
  {Colling}}]{Brucy2020}
{Brucy}, N., {Hennebelle}, P., {Bournaud}, F., \& {Colling}, C. 2020, \apjl,
  896, L34

\bibitem[{{Brucy} {et~al.}(2023){Brucy}, {Hennebelle}, {Colman}, \&
  {Iteanu}}]{Brucy2023}
{Brucy}, N., {Hennebelle}, P., {Colman}, T., \& {Iteanu}, S. 2023, arXiv
  e-prints, arXiv:2305.18012

\bibitem[{{Ciesla} {et~al.}(2023){Ciesla}, {G{\'o}mez-Guijarro}, {Buat},
  {Elbaz}, {Jin}, {B{\'e}thermin}, {Daddi}, {Franco}, {Inami}, {Magdis},
  {Magnelli}, \& {Xiao}}]{Ciesla2023}
{Ciesla}, L., {G{\'o}mez-Guijarro}, C., {Buat}, V., {et~al.} 2023, \aap, 672,
  A191

\bibitem[{{Cormier} {et~al.}(2014){Cormier}, {Madden}, {Lebouteiller}, {Hony},
  {Aalto}, {Costagliola}, {Hughes}, {R{\'e}my-Ruyer}, {Abel}, {Bayet},
  {Bigiel}, {Cannon}, {Cumming}, {Galametz}, {Galliano}, {Viti}, \&
  {Wu}}]{Cormier2014}
{Cormier}, D., {Madden}, S.~C., {Lebouteiller}, V., {et~al.} 2014, \aap, 564,
  A121

\bibitem[{{Daddi} {et~al.}(2010){Daddi}, {Elbaz}, {Walter}, {Bournaud},
  {Salmi}, {Carilli}, {Dannerbauer}, {Dickinson}, {Monaco}, \&
  {Riechers}}]{Daddi2010}
{Daddi}, E., {Elbaz}, D., {Walter}, F., {et~al.} 2010, \apjl, 714, L118

\bibitem[{{Dalgarno} \& {McCray}(1972)}]{dalgarno&mccray72}
{Dalgarno}, A. \& {McCray}, R.~A. 1972, \araa, 10, 375

\bibitem[{{de los Reyes} \& {Kennicutt}(2019)}]{delosReyes2019}
{de los Reyes}, M. A.~C. \& {Kennicutt}, Robert~C., J. 2019, \apj, 872, 16

\bibitem[{{Dekel} {et~al.}(2019){Dekel}, {Sarkar}, {Jiang}, {Bournaud},
  {Krumholz}, {Ceverino}, \& {Primack}}]{Dekel2019}
{Dekel}, A., {Sarkar}, K.~C., {Jiang}, F., {et~al.} 2019, \mnras, 488, 4753

\bibitem[{{Dubois} {et~al.}(2021){Dubois}, {Beckmann}, {Bournaud}, {Choi},
  {Devriendt}, {Jackson}, {Kaviraj}, {Kimm}, {Kraljic}, {Laigle}, {Martin},
  {Park}, {Peirani}, {Pichon}, {Volonteri}, \& {Yi}}]{Dubois2021}
{Dubois}, Y., {Beckmann}, R., {Bournaud}, F., {et~al.} 2021, \aap, 651, A109

\bibitem[{{Dubois} {et~al.}(2010){Dubois}, {Devriendt}, {Slyz}, \&
  {Teyssier}}]{duboisetal10}
{Dubois}, Y., {Devriendt}, J., {Slyz}, A., \& {Teyssier}, R. 2010, \mnras, 409,
  985

\bibitem[{{Dubois} {et~al.}(2012){Dubois}, {Devriendt}, {Slyz}, \&
  {Teyssier}}]{duboisetal12}
{Dubois}, Y., {Devriendt}, J., {Slyz}, A., \& {Teyssier}, R. 2012, \mnras, 420,
  2662

\bibitem[{{Dubois} {et~al.}(2014){Dubois}, {Pichon}, {Welker}, {Le Borgne},
  {Devriendt}, {Laigle}, {Codis}, {Pogosyan}, {Arnouts}, {Benabed}, {Bertin},
  {Blaizot}, {Bouchet}, {Cardoso}, \& {Colombi}}]{Dubois2014}
{Dubois}, Y., {Pichon}, C., {Welker}, C., {et~al.} 2014, \mnras, 444, 1453

\bibitem[{{Federrath} \& {Klessen}(2012)}]{federrath&klessen12}
{Federrath}, C. \& {Klessen}, R.~S. 2012, \apj, 761, 156

\bibitem[{{Federrath} {et~al.}(2008){Federrath}, {Klessen}, \&
  {Schmidt}}]{federrath2008}
{Federrath}, C., {Klessen}, R.~S., \& {Schmidt}, W. 2008, \apjl, 688, L79

\bibitem[{{Feldmann} {et~al.}(2012){Feldmann}, {Gnedin}, \&
  {Kravtsov}}]{Feldmann2012}
{Feldmann}, R., {Gnedin}, N.~Y., \& {Kravtsov}, A.~V. 2012, \apj, 758, 127

\bibitem[{{Filho} {et~al.}(2016){Filho}, {S{\'a}nchez Almeida}, {Amor{\'\i}n},
  {Mu{\~n}oz-Tu{\~n}{\'o}n}, {Elmegreen}, \& {Elmegreen}}]{Filho2016}
{Filho}, M.~E., {S{\'a}nchez Almeida}, J., {Amor{\'\i}n}, R., {et~al.} 2016,
  \apj, 820, 109

\bibitem[{{Freundlich} {et~al.}(2013){Freundlich}, {Combes}, {Tacconi},
  {Cooper}, {Genzel}, {Neri}, {Bolatto}, {Bournaud}, {Burkert}, {Cox}, {Davis},
  {F{\"o}rster Schreiber}, {Garcia-Burillo}, {Gracia-Carpio}, {Lutz}, {Naab},
  {Newman}, {Sternberg}, \& {Weiner}}]{Freundlich2013}
{Freundlich}, J., {Combes}, F., {Tacconi}, L.~J., {et~al.} 2013, \aap, 553,
  A130

\bibitem[{{Freundlich} {et~al.}(2019){Freundlich}, {Combes}, {Tacconi},
  {Genzel}, {Garcia-Burillo}, {Neri}, {Contini}, {Bolatto}, {Lilly},
  {Salom{\'e}}, {Bicalho}, {Boissier}, {Boone}, {Bouch{\'e}}, {Bournaud},
  {Burkert}, {Carollo}, {Cooper}, {Cox}, {Feruglio}, {F{\"o}rster Schreiber},
  {Juneau}, {Lippa}, {Lutz}, {Naab}, {Renzini}, {Saintonge}, {Sternberg},
  {Walter}, {Weiner}, {Wei{\ss}}, \& {Wuyts}}]{Freundlich2019}
{Freundlich}, J., {Combes}, F., {Tacconi}, L.~J., {et~al.} 2019, \aap, 622,
  A105

\bibitem[{{Genzel} {et~al.}(2010){Genzel}, {Tacconi}, {Gracia-Carpio},
  {Sternberg}, {Cooper}, {Shapiro}, {Bolatto}, {Bouch{\'e}}, {Bournaud},
  {Burkert}, {Combes}, {Comerford}, {Cox}, {Davis}, {F{\"o}rster Schreiber},
  {Garcia-Burillo}, {Lutz}, {Naab}, {Neri}, {Omont}, {Shapley}, \&
  {Weiner}}]{Genzel2010}
{Genzel}, R., {Tacconi}, L.~J., {Gracia-Carpio}, J., {et~al.} 2010, \mnras,
  407, 2091

\bibitem[{{Genzel} {et~al.}(2013){Genzel}, {Tacconi}, {Kurk}, {Wuyts},
  {Combes}, {Freundlich}, {Bolatto}, {Cooper}, {Neri}, {Nordon}, {Bournaud},
  {Burkert}, {Comerford}, {Cox}, {Davis}, {F{\"o}rster Schreiber},
  {Garc{\'\i}a-Burillo}, {Gracia-Carpio}, {Lutz}, {Naab}, {Newman},
  {Saintonge}, {Shapiro Griffin}, {Shapley}, {Sternberg}, \&
  {Weiner}}]{Genzel2013}
{Genzel}, R., {Tacconi}, L.~J., {Kurk}, J., {et~al.} 2013, \apj, 773, 68

\bibitem[{{G{\'o}mez-Guijarro} {et~al.}(2022){G{\'o}mez-Guijarro}, {Elbaz},
  {Xiao}, {Kokorev}, {Magdis}, {Magnelli}, {Daddi}, {Valentino}, {Sargent},
  {Dickinson}, {B{\'e}thermin}, {Franco}, {Pope}, {Kalita}, {Ciesla},
  {Demarco}, {Inami}, {Rujopakarn}, {Shu}, {Wang}, {Zhou}, {Alexander},
  {Bournaud}, {Chary}, {Ferguson}, {Finkelstein}, {Giavalisco}, {Iono},
  {Juneau}, {Kartaltepe}, {Lagache}, {Le Floc'h}, {Leiton}, {Leroy}, {Lin},
  {Motohara}, {Mullaney}, {Okumura}, {Pannella}, {Papovich}, \&
  {Treister}}]{Gomez2022}
{G{\'o}mez-Guijarro}, C., {Elbaz}, D., {Xiao}, M., {et~al.} 2022, \aap, 659,
  A196

\bibitem[{{Grisdale} {et~al.}(2022){Grisdale}, {Hogan}, {Rigopoulou}, {Thatte},
  {Pereira-Santaella}, {Devriendt}, {Slyz}, {Garc{\'\i}a-Bernete}, {Dubois},
  {Yi}, \& {Kraljic}}]{Grisdale2022}
{Grisdale}, K., {Hogan}, L., {Rigopoulou}, D., {et~al.} 2022, \mnras, 513, 3906

\bibitem[{{Haardt} \& {Madau}(1996)}]{haardt&madau96}
{Haardt}, F. \& {Madau}, P. 1996, \apj, 461, 20

\bibitem[{{Hennebelle} \& {Chabrier}(2011)}]{hennebelle&chabrier11}
{Hennebelle}, P. \& {Chabrier}, G. 2011, \apjl, 743, L29

\bibitem[{{Hogg} {et~al.}(2010){Hogg}, {Bovy}, \& {Lang}}]{Hogg2010}
{Hogg}, D.~W., {Bovy}, J., \& {Lang}, D. 2010, arXiv e-prints, arXiv:1008.4686

\bibitem[{{Hopkins} {et~al.}(2014){Hopkins}, {Kere{\v{s}}}, {O{\~n}orbe},
  {Faucher-Gigu{\`e}re}, {Quataert}, {Murray}, \& {Bullock}}]{Hopkins2014}
{Hopkins}, P.~F., {Kere{\v{s}}}, D., {O{\~n}orbe}, J., {et~al.} 2014, \mnras,
  445, 581

\bibitem[{{Hu} {et~al.}(2022){Hu}, {Smith}, {Teyssier}, {Bryan}, {Verbeke},
  {Emerick}, {Somerville}, {Burkhart}, {Li}, {Forbes}, \&
  {Starkenburg}}]{Hu2023}
{Hu}, C.-Y., {Smith}, M.~C., {Teyssier}, R., {et~al.} 2022, arXiv e-prints,
  arXiv:2208.10528

\bibitem[{{Hunter} {et~al.}(2021){Hunter}, {Elmegreen}, {Archer}, {Simpson}, \&
  {Cigan}}]{Hunter2021}
{Hunter}, D.~A., {Elmegreen}, B.~G., {Archer}, H., {Simpson}, C.~E., \&
  {Cigan}, P. 2021, \aj, 161, 175

\bibitem[{{Isobe} {et~al.}(1990){Isobe}, {Feigelson}, {Akritas}, \&
  {Babu}}]{Isobe1990}
{Isobe}, T., {Feigelson}, E.~D., {Akritas}, M.~G., \& {Babu}, G.~J. 1990, \apj,
  364, 104

\bibitem[{{Jackson} {et~al.}(2021{\natexlab{a}}){Jackson}, {Kaviraj}, {Martin},
  {Devriendt}, {Slyz}, {Silk}, {Dubois}, {Yi}, {Pichon}, {Volonteri}, {Choi},
  {Kimm}, {Kraljic}, \& {Peirani}}]{Jackson2021b}
{Jackson}, R.~A., {Kaviraj}, S., {Martin}, G., {et~al.} 2021{\natexlab{a}},
  \mnras, 502, 1785

\bibitem[{{Jackson} {et~al.}(2021{\natexlab{b}}){Jackson}, {Martin}, {Kaviraj},
  {Rams{\o}y}, {Devriendt}, {Sedgwick}, {Laigle}, {Choi}, {Beckmann},
  {Volonteri}, {Dubois}, {Pichon}, {Yi}, {Slyz}, {Kraljic}, {Kimm}, {Peirani},
  \& {Baldry}}]{Jackson2021a}
{Jackson}, R.~A., {Martin}, G., {Kaviraj}, S., {et~al.} 2021{\natexlab{b}},
  \mnras, 502, 4262

\bibitem[{{Kaviraj} {et~al.}(2017){Kaviraj}, {Laigle}, {Kimm}, {Devriendt},
  {Dubois}, {Pichon}, {Slyz}, {Chisari}, \& {Peirani}}]{Kaviraj2017}
{Kaviraj}, S., {Laigle}, C., {Kimm}, T., {et~al.} 2017, \mnras, 467, 4739

\bibitem[{{Kelly}(2007)}]{Kelly2007}
{Kelly}, B.~C. 2007, \apj, 665, 1489

\bibitem[{{Kennicutt}(1989)}]{Kennicutt1989}
{Kennicutt}, Robert~C., J. 1989, \apj, 344, 685

\bibitem[{{Kennicutt}(1998)}]{Kennicutt1998}
{Kennicutt}, Robert~C., J. 1998, \apj, 498, 541

\bibitem[{{Kennicutt} {et~al.}(2007){Kennicutt}, {Calzetti}, {Walter}, {Helou},
  {Hollenbach}, {Armus}, {Bendo}, {Dale}, {Draine}, {Engelbracht}, {Gordon},
  {Prescott}, {Regan}, {Thornley}, {Bot}, {Brinks}, {de Blok}, {de Mello},
  {Meyer}, {Moustakas}, {Murphy}, {Sheth}, \& {Smith}}]{Kennicutt2007}
{Kennicutt}, Robert~C., J., {Calzetti}, D., {Walter}, F., {et~al.} 2007, \apj,
  671, 333

\bibitem[{{Kennicutt} \& {De Los Reyes}(2021)}]{Kennicutt2021}
{Kennicutt}, Robert~C., J. \& {De Los Reyes}, M. A.~C. 2021, \apj, 908, 61

\bibitem[{{Kennicutt} \& {Evans}(2012)}]{Kennicutt2012}
{Kennicutt}, R.~C. \& {Evans}, N.~J. 2012, \araa, 50, 531

\bibitem[{{Kimm} \& {Cen}(2014)}]{kimm&cen14}
{Kimm}, T. \& {Cen}, R. 2014, \apj, 788, 121

\bibitem[{{Kimm} {et~al.}(2015){Kimm}, {Cen}, {Devriendt}, {Dubois}, \&
  {Slyz}}]{kimmetal15}
{Kimm}, T., {Cen}, R., {Devriendt}, J., {Dubois}, Y., \& {Slyz}, A. 2015,
  \mnras, 451, 2900

\bibitem[{{Kimm} {et~al.}(2017){Kimm}, {Katz}, {Haehnelt}, {Rosdahl},
  {Devriendt}, \& {Slyz}}]{kimmetal17}
{Kimm}, T., {Katz}, H., {Haehnelt}, M., {et~al.} 2017, \mnras, 466, 4826

\bibitem[{{Komatsu} {et~al.}(2011){Komatsu}, {Smith}, {Dunkley}, {Bennett},
  {Gold}, {Hinshaw}, {Jarosik}, {Larson}, {Nolta}, {Page}, {Spergel},
  {Halpern}, {Hill}, {Kogut}, {Limon}, {Meyer}, {Odegard}, {Tucker}, {Weiland},
  {Wollack}, \& {Wright}}]{komatsuetal11}
{Komatsu}, E., {Smith}, K.~M., {Dunkley}, J., {et~al.} 2011, \apjs, 192, 18

\bibitem[{{Kraljic} {et~al.}(2014){Kraljic}, {Renaud}, {Bournaud}, {Combes},
  {Elmegreen}, {Emsellem}, \& {Teyssier}}]{Kraljic2014}
{Kraljic}, K., {Renaud}, F., {Bournaud}, F., {et~al.} 2014, \apj, 784, 112

\bibitem[{{Kravtsov}(2003)}]{Kravtsov2003}
{Kravtsov}, A.~V. 2003, \apjl, 590, L1

\bibitem[{{Krumholz} \& {McKee}(2005)}]{krumholz&mckee05}
{Krumholz}, M.~R. \& {McKee}, C.~F. 2005, \apj, 630, 250

\bibitem[{{Leroy} {et~al.}(2008){Leroy}, {Walter}, {Brinks}, {Bigiel}, {de
  Blok}, {Madore}, \& {Thornley}}]{Leroy2008}
{Leroy}, A.~K., {Walter}, F., {Brinks}, E., {et~al.} 2008, \aj, 136, 2782

\bibitem[{{Leroy} {et~al.}(2013){Leroy}, {Walter}, {Sandstrom}, {Schruba},
  {Munoz-Mateos}, {Bigiel}, {Bolatto}, {Brinks}, {de Blok}, {Meidt}, {Rix},
  {Rosolowsky}, {Schinnerer}, {Schuster}, \& {Usero}}]{Leroy2013}
{Leroy}, A.~K., {Walter}, F., {Sandstrom}, K., {et~al.} 2013, \aj, 146, 19

\bibitem[{{Liu} {et~al.}(2015){Liu}, {Gao}, \& {Greve}}]{Liu2015}
{Liu}, L., {Gao}, Y., \& {Greve}, T.~R. 2015, \apj, 805, 31

\bibitem[{{Martin} {et~al.}(2021){Martin}, {Jackson}, {Kaviraj}, {Choi},
  {Devriendt}, {Dubois}, {Kimm}, {Kraljic}, {Peirani}, {Pichon}, {Volonteri},
  \& {Yi}}]{Martin2021}
{Martin}, G., {Jackson}, R.~A., {Kaviraj}, S., {et~al.} 2021, \mnras, 500, 4937

\bibitem[{{Matzner} \& {McKee}(2000)}]{matzner&mckee00}
{Matzner}, C.~D. \& {McKee}, C.~F. 2000, \apj, 545, 364

\bibitem[{{Narayanan} {et~al.}(2011){Narayanan}, {Krumholz}, {Ostriker}, \&
  {Hernquist}}]{Narayanan2011}
{Narayanan}, D., {Krumholz}, M., {Ostriker}, E.~C., \& {Hernquist}, L. 2011,
  \mnras, 418, 664

\bibitem[{{Onodera} {et~al.}(2010){Onodera}, {Kuno}, {Tosaki}, {Kohno},
  {Nakanishi}, {Sawada}, {Muraoka}, {Komugi}, {Miura}, {Kaneko}, {Hirota}, \&
  {Kawabe}}]{Onodera2010}
{Onodera}, S., {Kuno}, N., {Tosaki}, T., {et~al.} 2010, \apjl, 722, L127

\bibitem[{{Orr} {et~al.}(2018){Orr}, {Hayward}, {Hopkins}, {Chan},
  {Faucher-Gigu{\`e}re}, {Feldmann}, {Kere{\v{s}}}, {Murray}, \&
  {Quataert}}]{Orr2018}
{Orr}, M.~E., {Hayward}, C.~C., {Hopkins}, P.~F., {et~al.} 2018, \mnras, 478,
  3653

\bibitem[{{Padoan} \& {Nordlund}(2011)}]{padoan&nordlund11}
{Padoan}, P. \& {Nordlund}, {\r{A}}. 2011, \apj, 730, 40

\bibitem[{{Park} {et~al.}(2021){Park}, {Yi}, {Peirani}, {Pichon}, {Dubois},
  {Choi}, {Devriendt}, {Kaviraj}, {Kimm}, {Kraljic}, \& {Volonteri}}]{Park2021}
{Park}, M.~J., {Yi}, S.~K., {Peirani}, S., {et~al.} 2021, \apjs, 254, 2

\bibitem[{{Rawle} {et~al.}(2014){Rawle}, {Egami}, {Bussmann}, {Gurwell},
  {Ivison}, {Boone}, {Combes}, {Danielson}, {Rex}, {Richard}, {Smail},
  {Swinbank}, {Altieri}, {Blain}, {Clement}, {Dessauges-Zavadsky}, {Edge},
  {Fazio}, {Jones}, {Kneib}, {Omont}, {P{\'e}rez-Gonz{\'a}lez}, {Schaerer},
  {Valtchanov}, {van der Werf}, {Walth}, {Zamojski}, \& {Zemcov}}]{Rawle2014}
{Rawle}, T.~D., {Egami}, E., {Bussmann}, R.~S., {et~al.} 2014, \apj, 783, 59

\bibitem[{{Renaud} {et~al.}(2021{\natexlab{a}}){Renaud}, {Agertz}, {Andersson},
  {Read}, {Ryde}, {Bensby}, {Rey}, \& {Feuillet}}]{Renaud2021b}
{Renaud}, F., {Agertz}, O., {Andersson}, E.~P., {et~al.} 2021{\natexlab{a}},
  \mnras, 503, 5868

\bibitem[{{Renaud} {et~al.}(2021{\natexlab{b}}){Renaud}, {Agertz}, {Read},
  {Ryde}, {Andersson}, {Bensby}, {Rey}, \& {Feuillet}}]{Renaud2021}
{Renaud}, F., {Agertz}, O., {Read}, J.~I., {et~al.} 2021{\natexlab{b}}, \mnras,
  503, 5846

\bibitem[{{Renaud} {et~al.}(2019{\natexlab{a}}){Renaud}, {Bournaud}, {Agertz},
  {Kraljic}, {Schinnerer}, {Bolatto}, {Daddi}, \& {Hughes}}]{Renaud2019b}
{Renaud}, F., {Bournaud}, F., {Agertz}, O., {et~al.} 2019{\natexlab{a}}, \aap,
  625, A65

\bibitem[{{Renaud} {et~al.}(2019{\natexlab{b}}){Renaud}, {Bournaud}, {Daddi},
  \& {Wei{\ss}}}]{Renaud2019a}
{Renaud}, F., {Bournaud}, F., {Daddi}, E., \& {Wei{\ss}}, A.
  2019{\natexlab{b}}, \aap, 621, A104

\bibitem[{{Renaud} {et~al.}(2014){Renaud}, {Bournaud}, {Kraljic}, \&
  {Duc}}]{Renaud2014}
{Renaud}, F., {Bournaud}, F., {Kraljic}, K., \& {Duc}, P.~A. 2014, \mnras, 442,
  L33

\bibitem[{{Renaud} {et~al.}(2012){Renaud}, {Kraljic}, \&
  {Bournaud}}]{Renaud2012}
{Renaud}, F., {Kraljic}, K., \& {Bournaud}, F. 2012, \apjl, 760, L16

\bibitem[{{Renaud} {et~al.}(2022){Renaud}, {Segovia Otero}, \&
  {Agertz}}]{Renaud2022}
{Renaud}, F., {Segovia Otero}, {\'A}., \& {Agertz}, O. 2022, \mnras, 516, 4922

\bibitem[{{Rosdahl} \& {Blaizot}(2012)}]{rosdahl&blaizot12}
{Rosdahl}, J. \& {Blaizot}, J. 2012, \mnras, 423, 344

\bibitem[{{Roychowdhury} {et~al.}(2017){Roychowdhury}, {Chengalur}, \&
  {Shi}}]{Roychowdhury2017}
{Roychowdhury}, S., {Chengalur}, J.~N., \& {Shi}, Y. 2017, \aap, 608, A24

\bibitem[{{Salmi} {et~al.}(2012){Salmi}, {Daddi}, {Elbaz}, {Sargent},
  {Dickinson}, {Renzini}, {Bethermin}, \& {Le Borgne}}]{Salmi2012}
{Salmi}, F., {Daddi}, E., {Elbaz}, D., {et~al.} 2012, \apjl, 754, L14

\bibitem[{{Schmidt}(1959)}]{Schmidt1959}
{Schmidt}, M. 1959, \apj, 129, 243

\bibitem[{{Schruba} {et~al.}(2011){Schruba}, {Leroy}, {Walter}, {Bigiel},
  {Brinks}, {de Blok}, {Dumas}, {Kramer}, {Rosolowsky}, {Sandstrom},
  {Schuster}, {Usero}, {Weiss}, \& {Wiesemeyer}}]{Schruba2011}
{Schruba}, A., {Leroy}, A.~K., {Walter}, F., {et~al.} 2011, \aj, 142, 37

\bibitem[{{Segovia Otero} {et~al.}(2022){Segovia Otero}, {Renaud}, \&
  {Agertz}}]{Segovia2022}
{Segovia Otero}, {\'A}., {Renaud}, F., \& {Agertz}, O. 2022, \mnras, 516, 2272

\bibitem[{{Semenov} {et~al.}(2019){Semenov}, {Kravtsov}, \&
  {Gnedin}}]{Semenov2019}
{Semenov}, V.~A., {Kravtsov}, A.~V., \& {Gnedin}, N.~Y. 2019, \apj, 870, 79

\bibitem[{{Sharon} {et~al.}(2013){Sharon}, {Baker}, {Harris}, \&
  {Thomson}}]{Sharon2013}
{Sharon}, C.~E., {Baker}, A.~J., {Harris}, A.~I., \& {Thomson}, A.~P. 2013,
  \apj, 765, 6

\bibitem[{{Shetty} {et~al.}(2013){Shetty}, {Kelly}, \& {Bigiel}}]{Shetty2013}
{Shetty}, R., {Kelly}, B.~C., \& {Bigiel}, F. 2013, \mnras, 430, 288

\bibitem[{{Sun} {et~al.}(2023){Sun}, {Leroy}, {Ostriker}, {Meidt},
  {Rosolowsky}, {Schinnerer}, {Wilson}, {Utomo}, {Belfiore}, {Blanc},
  {Emsellem}, {Faesi}, {Groves}, {Hughes}, {Koch}, {Kreckel}, {Liu}, {Pan},
  {Pety}, {Querejeta}, {Razza}, {Saito}, {Sardone}, {Usero}, {Williams},
  {Bigiel}, {Bolatto}, {Chevance}, {Dale}, {Gensior}, {Glover}, {Grasha},
  {Henshaw}, {Jim{\'e}nez-Donaire}, {Klessen}, {Kruijssen}, {Murphy},
  {Neumann}, {Teng}, \& {Thilker}}]{Sun2023}
{Sun}, J., {Leroy}, A.~K., {Ostriker}, E.~C., {et~al.} 2023, \apjl, 945, L19

\bibitem[{{Sutherland} \& {Dopita}(1993)}]{sutherland&dopita93}
{Sutherland}, R.~S. \& {Dopita}, M.~A. 1993, \apjs, 88, 253

\bibitem[{{Tacconi} {et~al.}(2010){Tacconi}, {Genzel}, {Neri}, {Cox}, {Cooper},
  {Shapiro}, {Bolatto}, {Bouch{\'e}}, {Bournaud}, {Burkert}, {Combes},
  {Comerford}, {Davis}, {F{\"o}rster Schreiber}, {Garcia-Burillo},
  {Gracia-Carpio}, {Lutz}, {Naab}, {Omont}, {Shapley}, {Sternberg}, \&
  {Weiner}}]{Tacconi2010}
{Tacconi}, L.~J., {Genzel}, R., {Neri}, R., {et~al.} 2010, \nat, 463, 781

\bibitem[{{Tacconi} {et~al.}(2018){Tacconi}, {Genzel}, {Saintonge}, {Combes},
  {Garc{\'\i}a-Burillo}, {Neri}, {Bolatto}, {Contini}, {F{\"o}rster Schreiber},
  {Lilly}, {Lutz}, {Wuyts}, {Accurso}, {Boissier}, {Boone}, {Bouch{\'e}},
  {Bournaud}, {Burkert}, {Carollo}, {Cooper}, {Cox}, {Feruglio}, {Freundlich},
  {Herrera-Camus}, {Juneau}, {Lippa}, {Naab}, {Renzini}, {Salome}, {Sternberg},
  {Tadaki}, {{\"U}bler}, {Walter}, {Weiner}, \& {Weiss}}]{Tacconi2018}
{Tacconi}, L.~J., {Genzel}, R., {Saintonge}, A., {et~al.} 2018, \apj, 853, 179

\bibitem[{{Tacconi} {et~al.}(2013){Tacconi}, {Neri}, {Genzel}, {Combes},
  {Bolatto}, {Cooper}, {Wuyts}, {Bournaud}, {Burkert}, {Comerford}, {Cox},
  {Davis}, {F{\"o}rster Schreiber}, {Garc{\'\i}a-Burillo}, {Gracia-Carpio},
  {Lutz}, {Naab}, {Newman}, {Omont}, {Saintonge}, {Shapiro Griffin}, {Shapley},
  {Sternberg}, \& {Weiner}}]{Tacconi2013}
{Tacconi}, L.~J., {Neri}, R., {Genzel}, R., {et~al.} 2013, \apj, 768, 74

\bibitem[{{Teng} {et~al.}(2023){Teng}, {Sandstrom}, {Sun}, {Gong}, {Bolatto},
  {Chiang}, {Leroy}, {Usero}, {Glover}, {Klessen}, {Liu}, {Querejeta},
  {Schinnerer}, {Bigiel}, {Cao}, {Chevance}, {Eibensteiner}, {Grasha},
  {Israel}, {Murphy}, {Neumann}, {Pan}, {Pinna}, {Sormani}, {Smith}, {Walter},
  \& {Williams}}]{Teng2023}
{Teng}, Y.-H., {Sandstrom}, K.~M., {Sun}, J., {et~al.} 2023, arXiv e-prints,
  arXiv:2304.04732

\bibitem[{{Teyssier}(2002)}]{teyssier02}
{Teyssier}, R. 2002, \aap, 385, 337

\bibitem[{{Teyssier} {et~al.}(2011){Teyssier}, {Moore}, {Martizzi}, {Dubois},
  \& {Mayer}}]{teyssieretal11}
{Teyssier}, R., {Moore}, B., {Martizzi}, D., {Dubois}, Y., \& {Mayer}, L. 2011,
  \mnras, 414, 195

\bibitem[{{Thornton} {et~al.}(1998){Thornton}, {Gaudlitz}, {Janka}, \&
  {Steinmetz}}]{Thornton1998}
{Thornton}, K., {Gaudlitz}, M., {Janka}, H.~T., \& {Steinmetz}, M. 1998, \apj,
  500, 95

\bibitem[{{Trebitsch} {et~al.}(2017){Trebitsch}, {Blaizot}, {Rosdahl},
  {Devriendt}, \& {Slyz}}]{trebitschetal17}
{Trebitsch}, M., {Blaizot}, J., {Rosdahl}, J., {Devriendt}, J., \& {Slyz}, A.
  2017, \mnras, 470, 224

\bibitem[{{Trebitsch} {et~al.}(2021){Trebitsch}, {Dubois}, {Volonteri},
  {Pfister}, {Cadiou}, {Katz}, {Rosdahl}, {Kimm}, {Pichon}, {Beckmann},
  {Devriendt}, \& {Slyz}}]{trebitschetal20}
{Trebitsch}, M., {Dubois}, Y., {Volonteri}, M., {et~al.} 2021, \aap, 653, A154

\bibitem[{{Vazquez-Semadeni}(1994)}]{Vazquez1994}
{Vazquez-Semadeni}, E. 1994, \apj, 423, 681

\bibitem[{{Volonteri} {et~al.}(2020){Volonteri}, {Pfister}, {Beckmann},
  {Dubois}, {Colpi}, {Conselice}, {Dotti}, {Martin}, {Jackson}, {Kraljic},
  {Pichon}, {Trebitsch}, {Yi}, {Devriendt}, \& {Peirani}}]{Volonteri2020}
{Volonteri}, M., {Pfister}, H., {Beckmann}, R.~S., {et~al.} 2020, \mnras, 498,
  2219

\end{thebibliography}
 
\begin{appendix} 

\section{Galaxy distribution in  KS plane}
\subsection{Dependence on the total gas}

\begin{figure*}
\centering\includegraphics[width=0.98\textwidth]{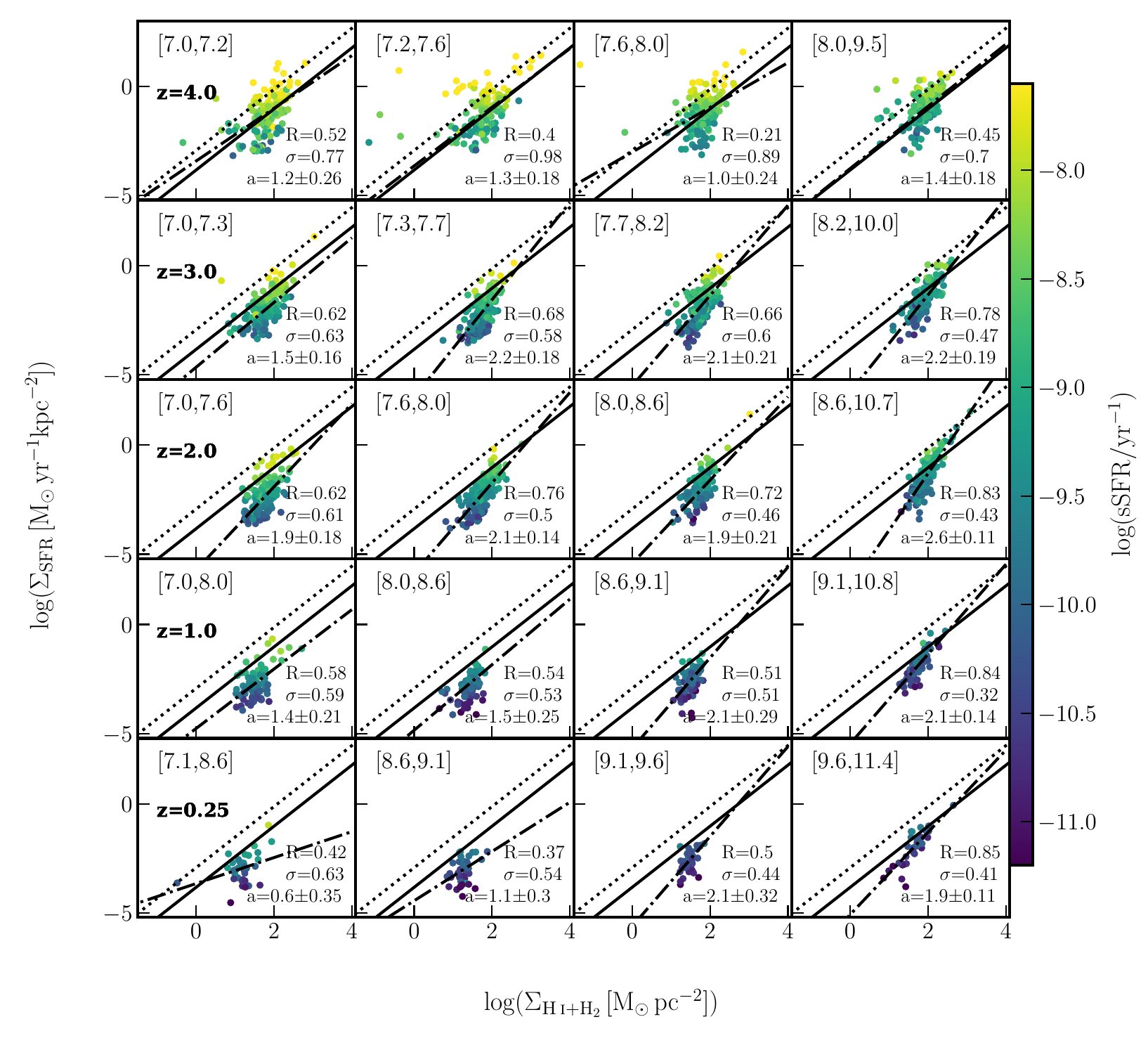}
\caption{Same as \protect Fig.~\ref{fig:KS_SFR10_gasSF_ssfr}, but for the neutral gas surface density \sigmagastot.
}
\label{fig:KS_SFR10_gastot_ssfr}
\end{figure*}

Figure~\ref{fig:KS_SFR10_gastot_ssfr} shows the distribution of galaxies in the \sigmagastot-\sigmasfr parameter space at different redshifts (rows) and in four equally-populated stellar mass quartiles at each redshift (columns). 
Trends with the \ssfr (given by the colour coding) seen when considering the molecular gas (see Fig.~\ref{fig:KS_SFR10_gasSF_ssfr}) persist. At each redshift and each stellar mass bin, the \ssfr
of galaxies strongly correlated with \sigmasfr, and 
regardless of the stellar mass, the distributions of
galaxies move within the KS parameter space towards lower values of \sigmagastot and \sigmasfr with decreasing redshift.

\subsection{Molecular and total gas fractions}

\begin{figure*}
\centering\includegraphics[width=0.98\textwidth]{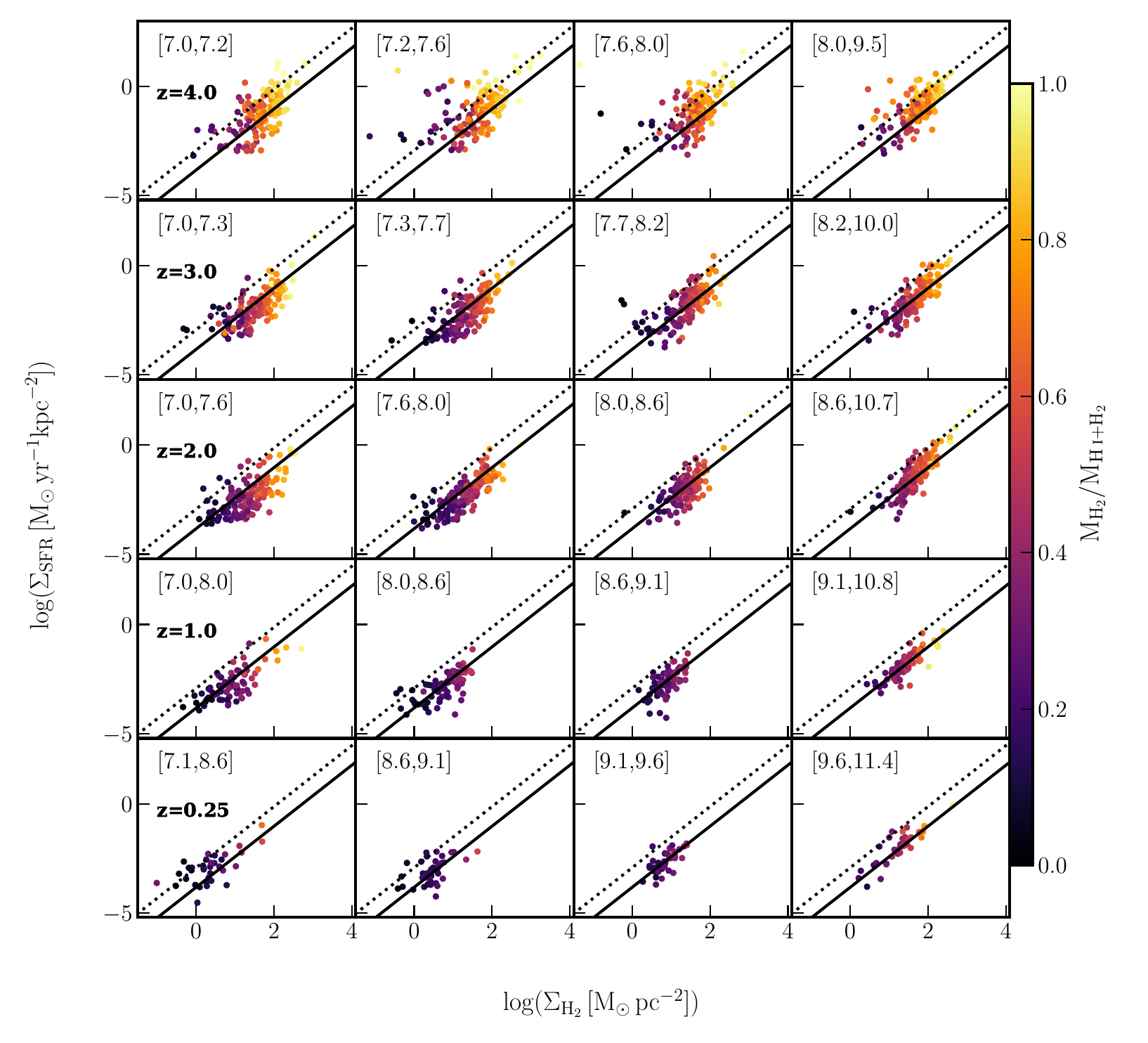}
\caption{Same as \protect Fig.~\ref{fig:KS_SFR10_gasSF_ssfr}, but colour-coded by the molecular over cold gas fraction.
}
\label{fig:KS_SFR10_gasSF_fgas_SF_overtot}
\end{figure*}

Figure~\ref{fig:KS_SFR10_gasSF_fgas_SF_overtot} shows the molecular gas fraction, defined as the fraction of \hmol mass over the neutral gas, i.e. M$_{\rm{H}_2}$/(M$_{\textsc{H\,i}} + \rm{M}_{\rm{H}_2}$), and its evolution within the KS plane with the cosmic time for different stellar mass bins.

\begin{figure*}
\centering\includegraphics[width=0.98\textwidth]{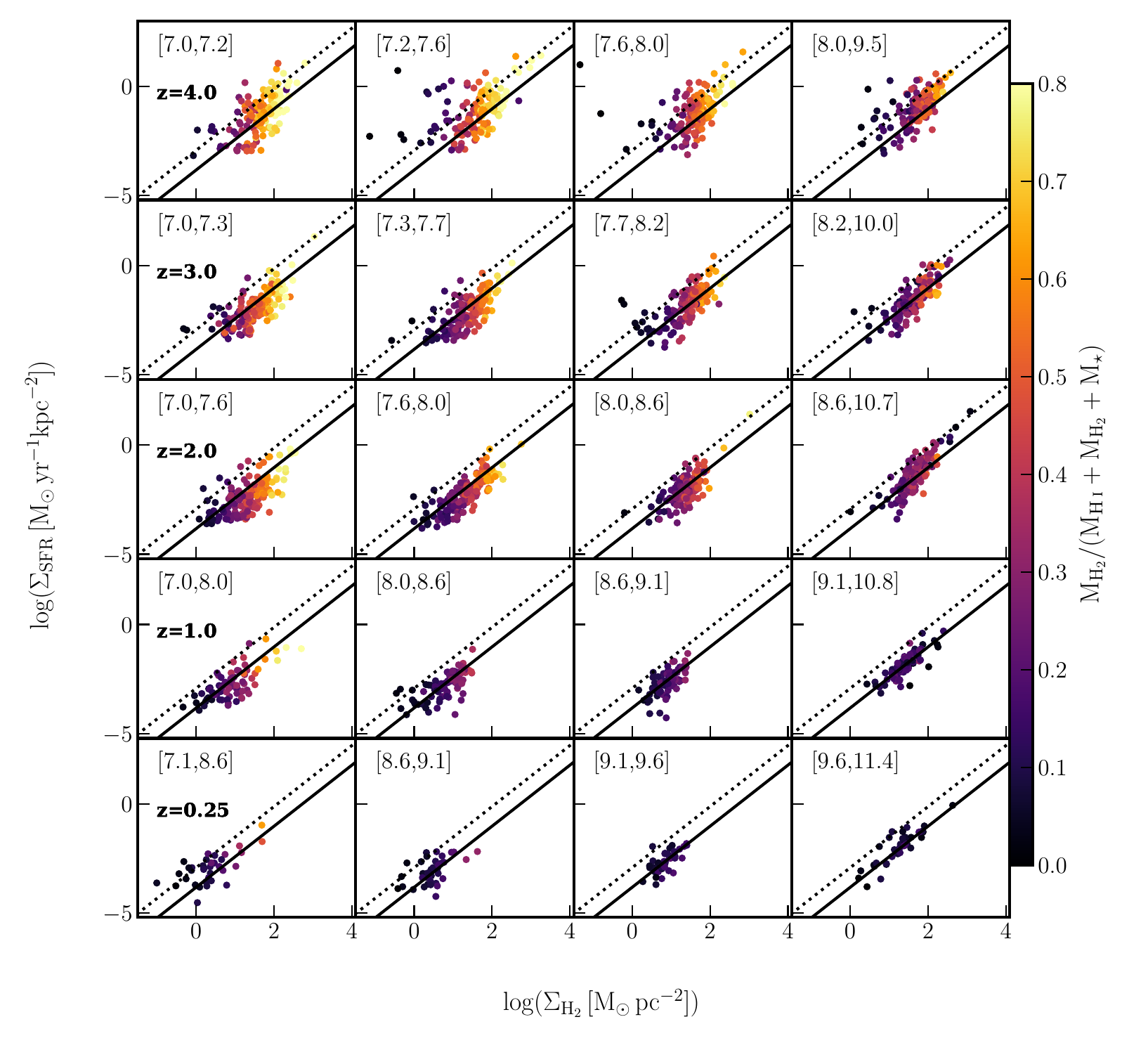}
\caption{Same as \protect Fig.~\ref{fig:KS_SFR10_gasSF_ssfr}, but colour-coded by the baryonic molecular gas fraction. 
}
\label{fig:KS_SFR10_gasSF_fgasSF}
\end{figure*}

Figure~\ref{fig:KS_SFR10_gasSF_fgasSF} shows the molecular gas content of galaxies defined in terms of baryonic fraction, i.e.  M$_{\rm{H}_2}/(\rm{M}_{\textsc{H\,i}} + \rm{M}_{\rm{H}_2} + \rm{M}_\star)$, as a function of cosmic time and stellar mass.

\begin{figure*}
\centering\includegraphics[width=0.98\textwidth]{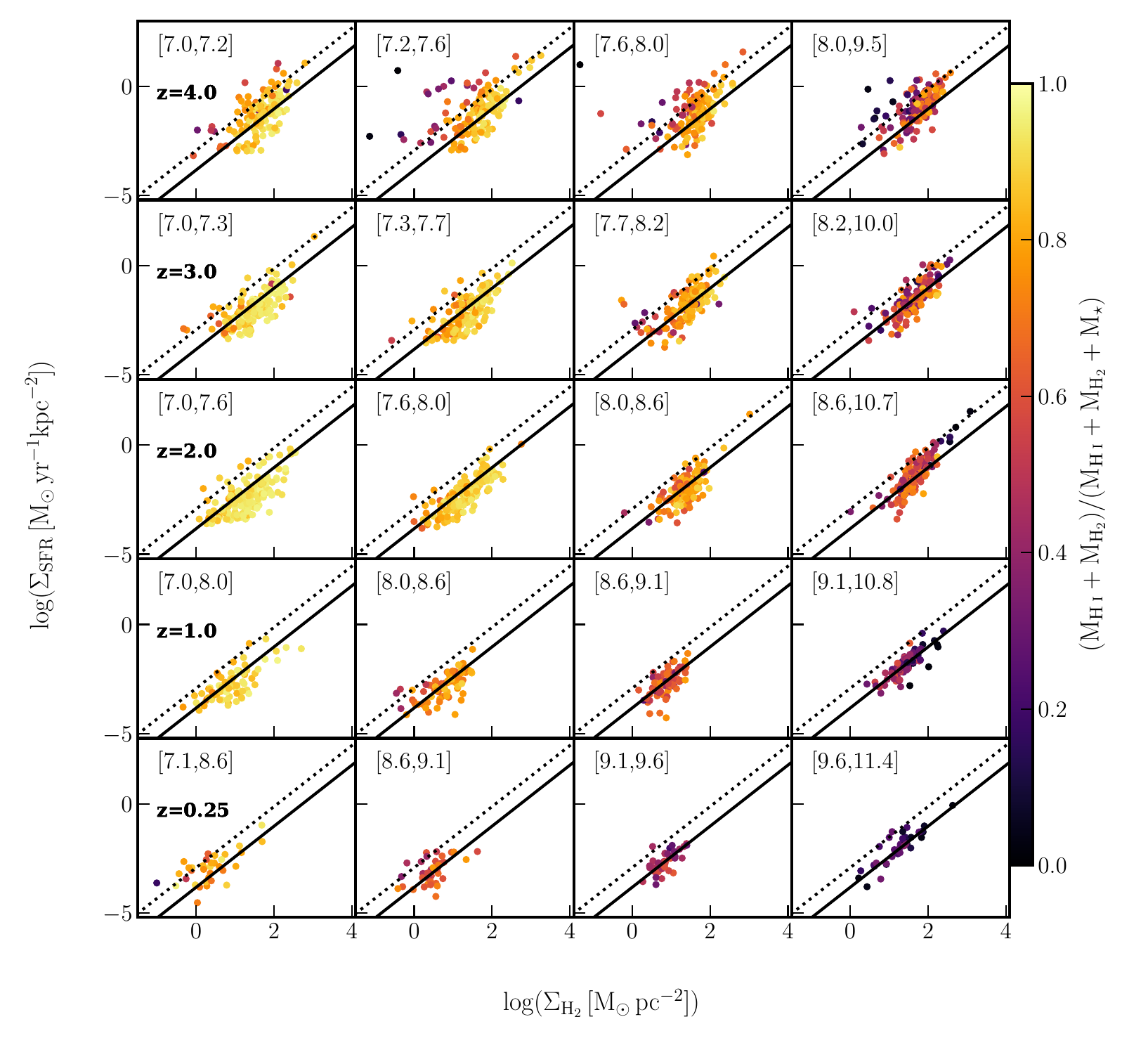}
\caption{Same as \protect Fig.~\ref{fig:KS_SFR10_gasSF_ssfr}, but colour-coded by the neutral gas fraction.
}
\label{fig:KS_SFR10_gasSF_fgastot}
\end{figure*}

Figure~\ref{fig:KS_SFR10_gasSF_fgastot} shows the neutral baryonic gas fraction, i.e. (M$_{\rm \textsc{H\,i}} + \rm{M}_{\rm{H}_2})/(\rm{M}_{\textsc{H\,i}} + \rm{M}_{\rm{H}_2} + \rm{M}_\star)$, at different redshifts and in different stellar mass bins.

\subsection{Turbulence}

\begin{figure*}
\centering\includegraphics[width=0.98\textwidth]{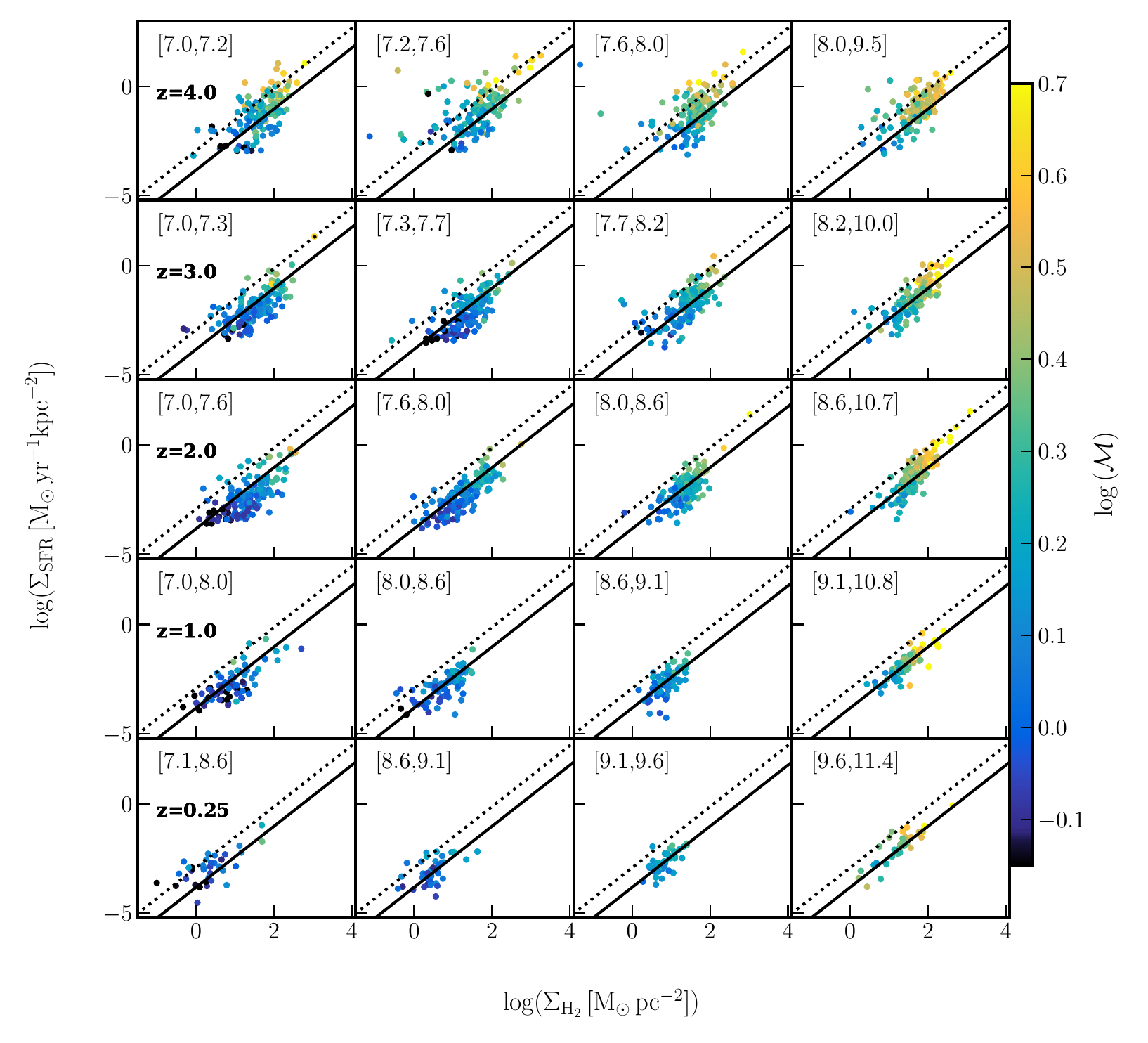}
\caption{Same as \protect Fig.~\ref{fig:KS_SFR10_gasSF_ssfr}, but colour-coded by the Mach number.
}
\label{fig:KS_SFR10_gastot_Mach}
\end{figure*}

Figure~\ref{fig:KS_SFR10_gastot_Mach} shows 
the distributions of galaxies in the \sigmagasSF-\sigmasfr parameter space as a function of Mach number \mach. 
At all redshifts and in all stellar mass bins \mach monotonically increases with increasing \sigmagasSF and \sigmasfr. In addition, at fixed \sigmagasSF, \sigmasfr correlates with \mach at all redshifts, driving the offset of the most turbulent
galaxies above the observed KS relation onto the starburst regime (dotted line) and beyond, as visible at high redshift. 

At $z<2$, this regime is very sparsely populated. 
At fixed stellar mass, galaxies at high redshifts are
more turbulent than their low redshift counterparts, and at fixed redshift, more massive galaxies tend to be more turbulent than their lower mass counterparts.

\begin{figure*}
\centering\includegraphics[width=0.98\textwidth]{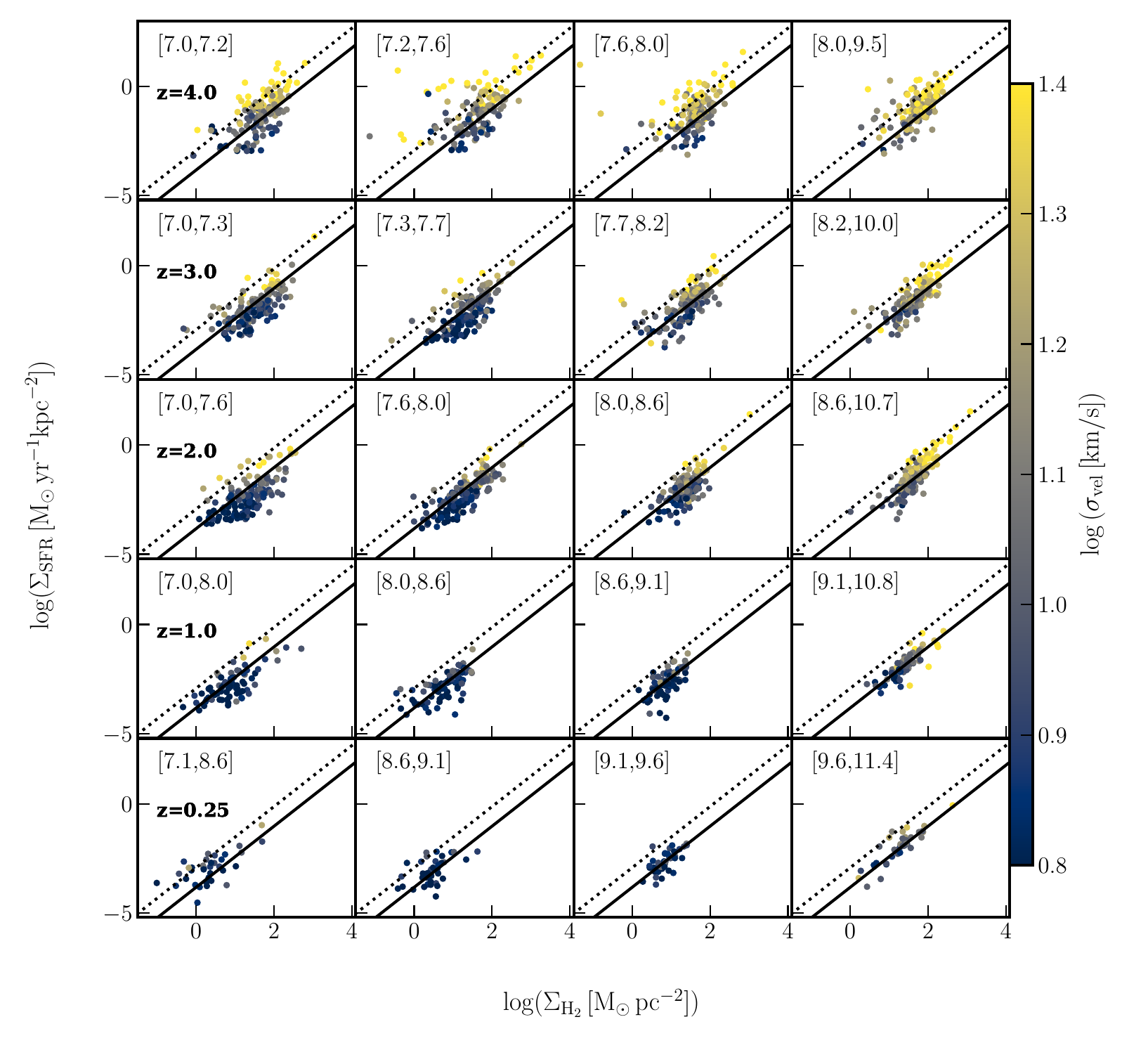}
\caption{Same as \protect Fig.~\ref{fig:KS_SFR10_gasSF_ssfr}, but colour-coded by the gas velocity dispersion.
}
\label{fig:KS_SFR10_gasSF_vsig}
\end{figure*}

\begin{figure*}
\centering\includegraphics[width=0.98\textwidth]{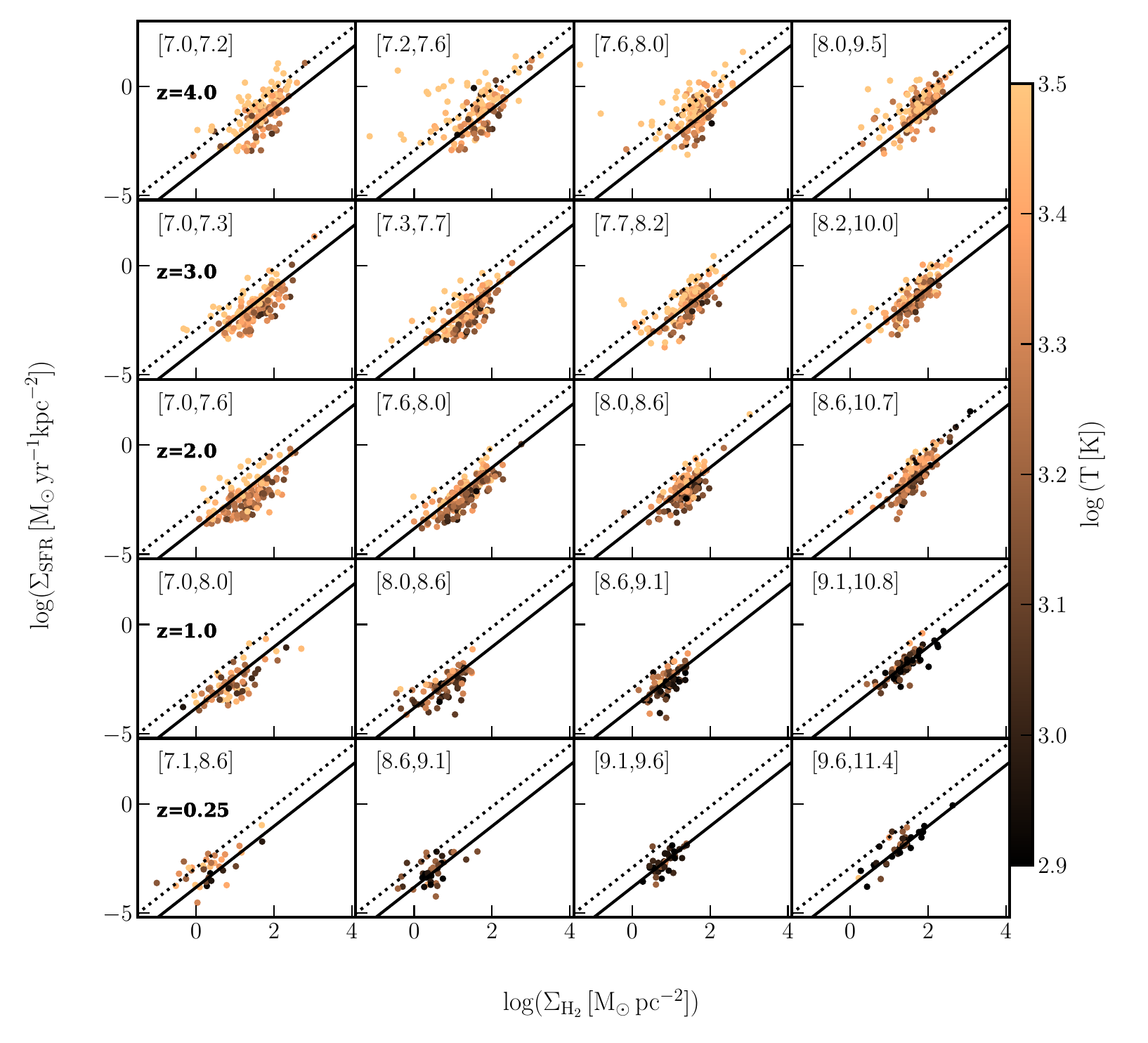}
\caption{Same as \protect Fig.~\ref{fig:KS_SFR10_gasSF_ssfr}, but colour-coded by the gas temperature.
}
\label{fig:KS_SFR10_gasSF_T}
\end{figure*}

Figures~\ref{fig:KS_SFR10_gasSF_vsig} and \ref{fig:KS_SFR10_gasSF_T} correspond to the distribution of galaxies in the \sigmagasSF-\sigmasfr parameter space color-coded by their gas velocity dispersion and temperature, respectively. 

\section{Fitting methods}
\label{sec:bayesian}

We fit the distribution of galaxies within the $\log \Sigma_{\rm gas}-\log \Sigma_{\rm SFR}$ plane with the linear relation $\log \Sigma_{\rm SFR} = a(\log \Sigma_{\rm gas}) + b$, with the best-fit values for the slope $a$ and intercept $b$ given in each panel. We compared three different fitting methods, largely used and advocated in the literature. These are the ordinary least square (OLS) technique, OLS bisector (OLS-bis) technique \citep{Isobe1990}, and Bayesian linear regression (BLR) with the Student's t-distribution\footnote{We use the probabilistic programming package for Python \pymc, available at 
\href{https://docs.pymc.io/en/v3/index.html}{https://docs.pymc.io/en/v3/index.html}.
} for the likelihood to minimise the impact of outliers. Normal likelihood gives the data used in this work identical results as OLS. The same is true for the hierarchical Bayesian model \textsl{linmix} \citep{Kelly2007} that we have tested as well\footnote{We use the Python package implementing the linmix algorithm available on github at \href{https://github.com/jmeyers314/linmix}{https://github.com/jmeyers314/linmix}.}.
Overall, comparing the three above-mentioned methods, for the data used in this work, we conclude that:
\begin{itemize}
    \item For the fits involving the entire population of galaxies (e.g. Fig.~\ref{fig:KS_ssfr_gas_SF_tot}), the derived slopes show the following hierarchy $a_{\rm{OLS}} \leq a_{\rm{BLR}} < a_{\rm{OLS-bis}}$ at all redshifts (with $a_{\rm{OLS}}$ and $a_{\rm{BLR}}$ being often consistent within the error-bars) and follows the same trend with redshift. The dispersion around the best fit (i.e. the rms in $\log(\sigmasfr)$) follows the same trend.  
    This is true for both \sigmagasSF- and \sigmagastot-\sigmasfr fits.
    \item For the fits at different mass bins and redshifts (e.g. Fig.~\ref{fig:KS_SFR10_gasSF_ssfr}) the same hierarchy is followed by slopes with $a_{\rm{OLS}}$ and $a_{\rm{BLR}}$ being most of the time within uncertainties of each other, for both \sigmagasSF and \sigmagastot. The same applies to the dispersion around the best fit.
\end{itemize}
All three explored fitting methods provide qualitatively similar trends for the slopes and dispersions around the best fit as a function of redshift and stellar mass. 
There are however notable quantitative differences, with slopes being systematically higher for the bisector OLS method compared to the two other methods. Similarly, the dispersion around the best fit is always larger for the bisector OLS method compared to the standard OLS and Bayesian methods.

\subsection{Distribution of galaxies in the KS plane}

\begin{figure*}
\centering\includegraphics[width=0.98\textwidth]{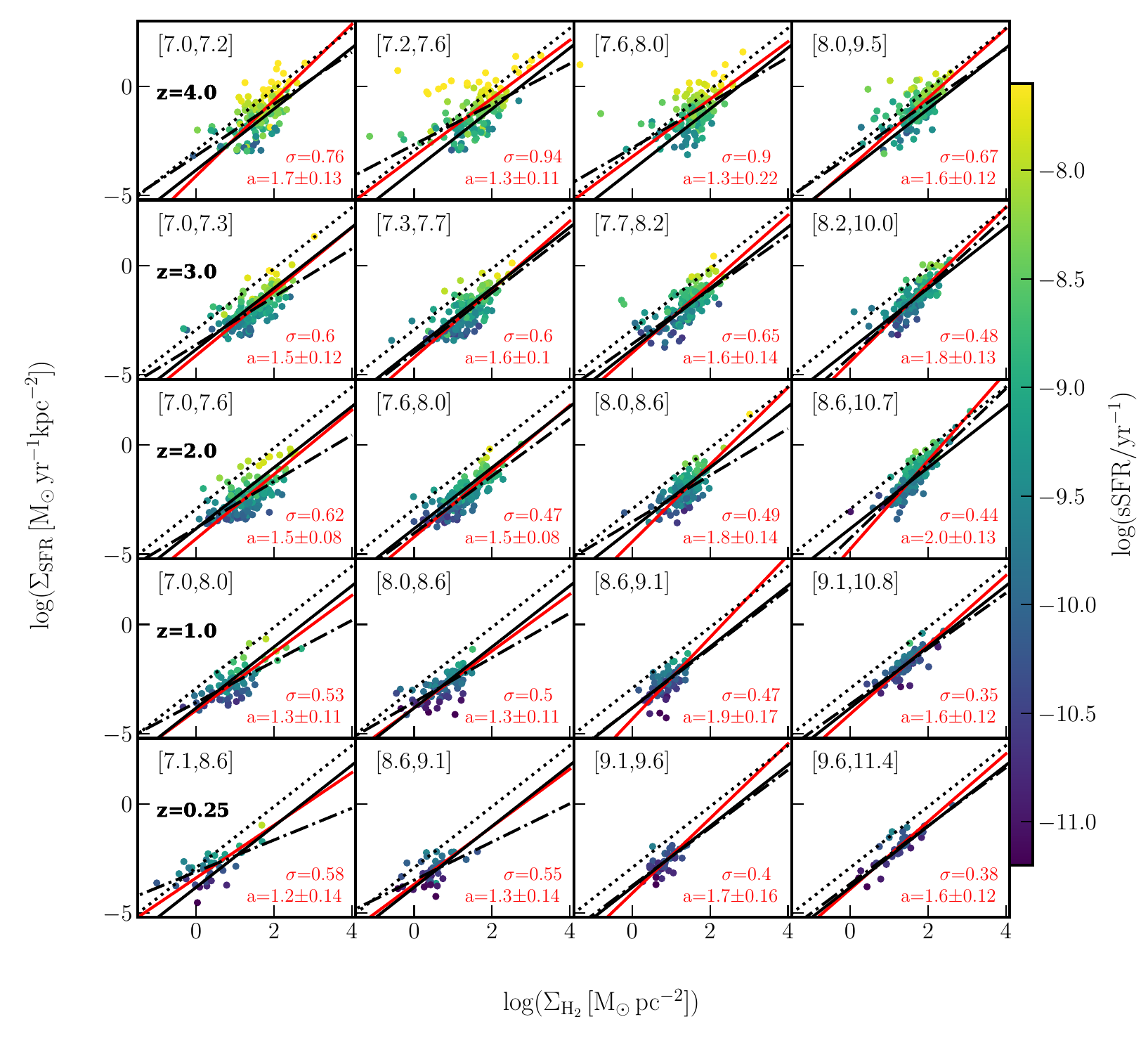}
\caption{Same as Fig.~\ref{fig:KS_SFR10_gasSF_ssfr}, but showing in addition, OLS bisector fits for comparison (red solid lines). The corresponding best-fit values for slope $a$ and the standard deviation of residuals $\sigma$ are shown in the bottom right corners of each panel.
}
\label{fig:KS_SFR10_gasSF_bisec_ssfr}
\end{figure*}

Figure~\ref{fig:KS_SFR10_gasSF_bisec_ssfr} shows the distribution of galaxies in the \sigmasfr-\sigmagasSF plane, as in Fig.~\ref{fig:KS_SFR10_gasSF_ssfr}, but reporting the results of the OLS bisector fits for comparison (red solid lines) to the Bayesian fit (black dash-dotted lines). The corresponding best-fit values for the OLS fit slope $a$ and the standard deviation of residuals $\sigma$ are shown in the bottom right corners of each panel.

\begin{figure*}
\centering\includegraphics[width=0.98\textwidth]{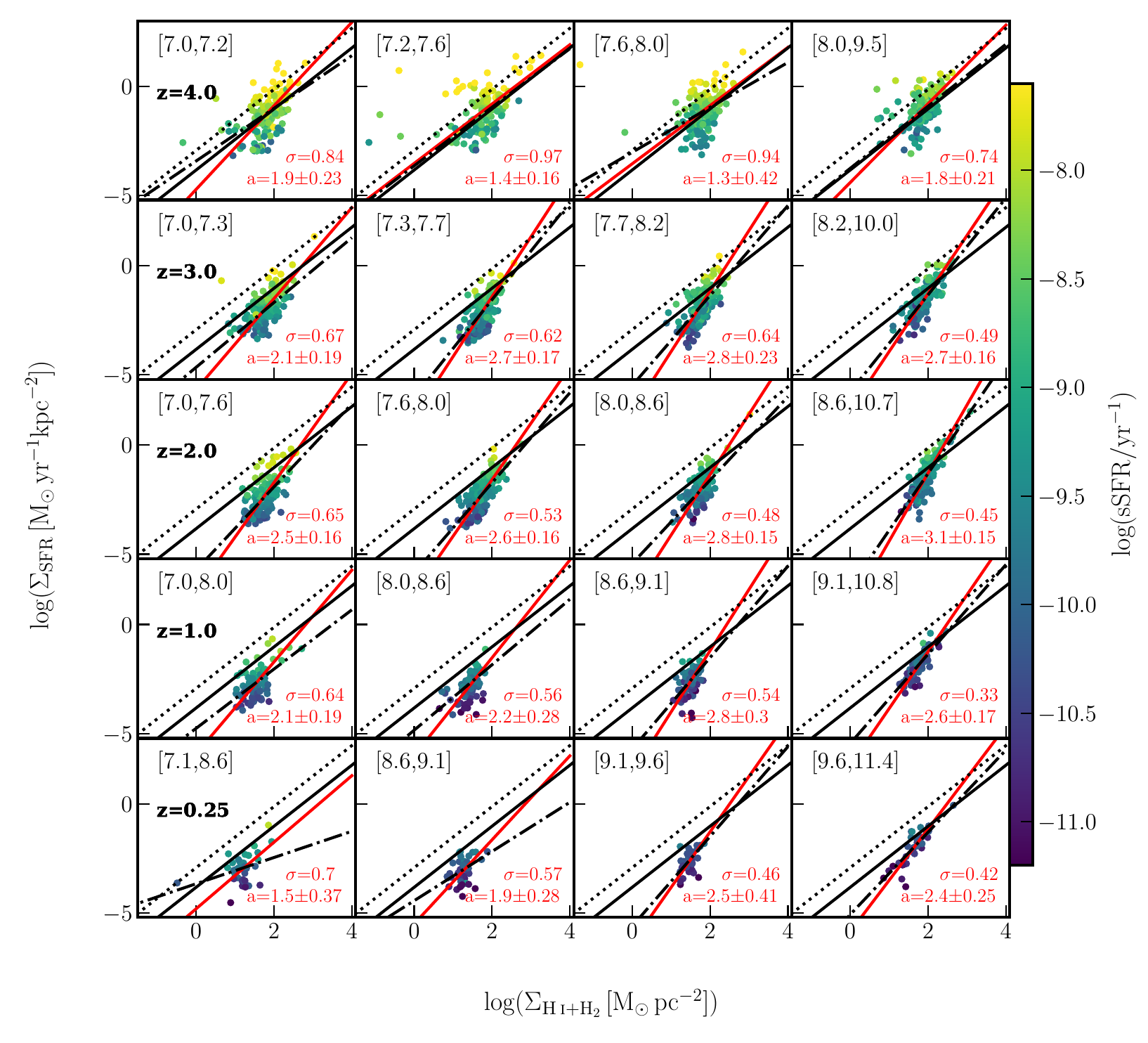}
\caption{Same as Fig.~\ref{fig:KS_SFR10_gasSF_bisec_ssfr}, but for the neutral gas surface density \sigmagastot.
}
\label{fig:KS_SFR10_gastot_bisec_ssfr}
\end{figure*}

Figure~\ref{fig:KS_SFR10_gastot_bisec_ssfr} is the same as Fig.~\ref{fig:KS_SFR10_gasSF_bisec_ssfr}, but showing instead the distribution of galaxies in the \sigmasfr-\sigmagastot parameter space.

\subsection{Integrated KS}

\begin{figure*}
\centering\includegraphics[width=0.9\textwidth]{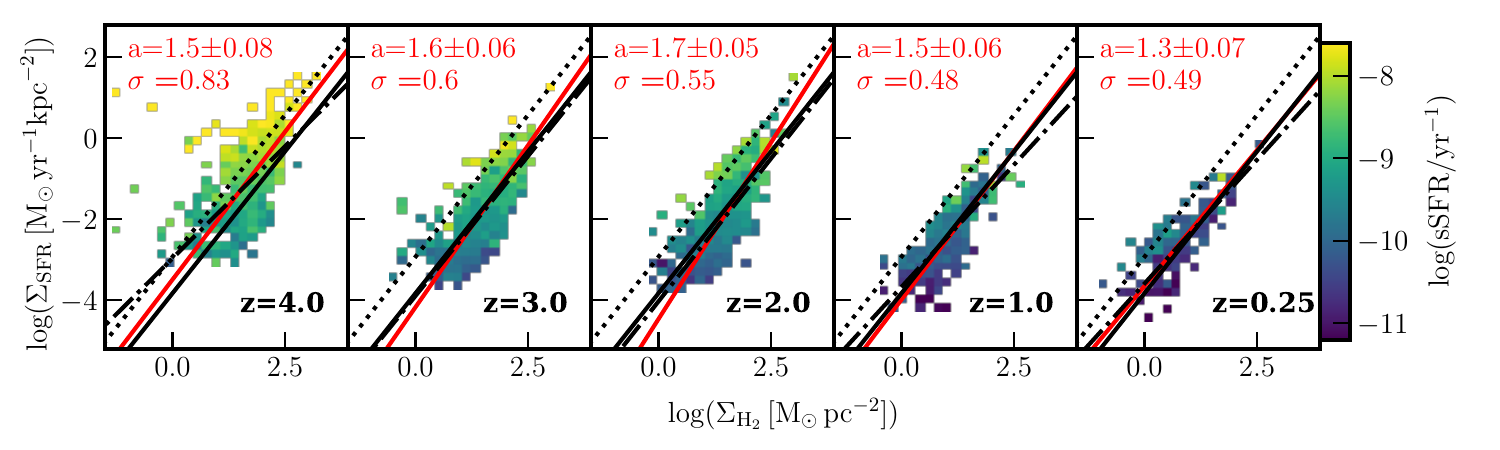}\\[-0.1cm]
\centering\includegraphics[width=0.9\textwidth]{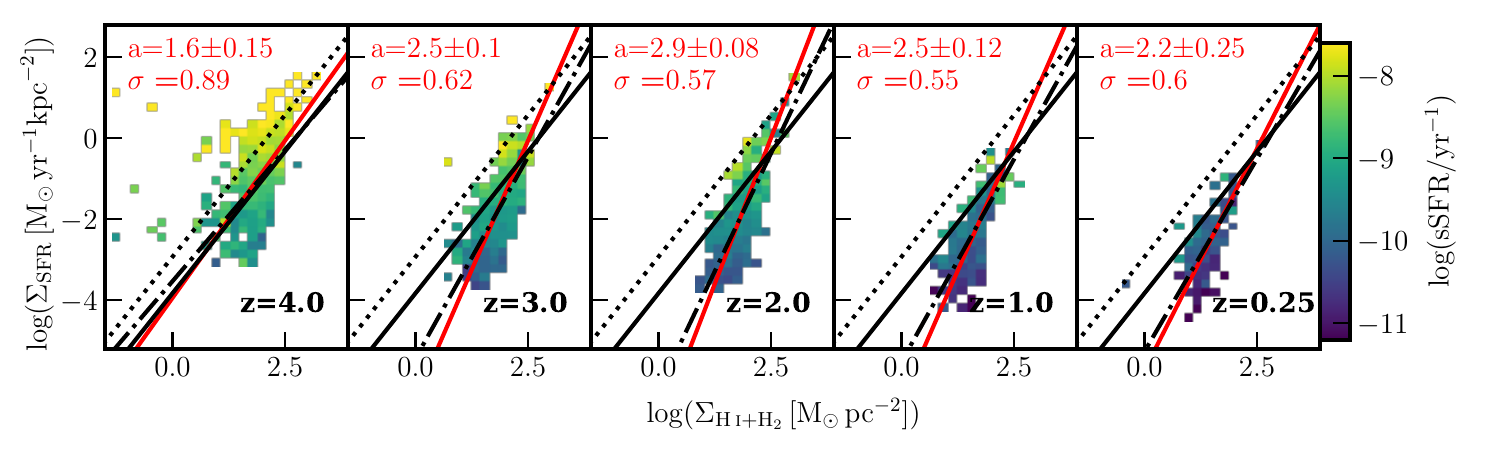}
\caption{Same as Fig.~\ref{fig:KS_ssfr_gas_SF_tot}, 
but showing in addition, OLS bisector fits for comparison (red solid lines). The corresponding best-fit values for slope $a$ and the standard deviation of residuals $\sigma$ are shown in the bottom right corners of each panel.
}
\label{fig:KS_ssfr_gas_SF_tot_bayes}
\end{figure*}

Figure~\ref{fig:KS_ssfr_gas_SF_tot_bayes} is equivalent to Fig.~\ref{fig:KS_ssfr_gas_SF_tot}, but showing the results of the OLS bisector fit (solid red lines), together with its best-fit values that can be compared to the Bayesian fit (black dash-dotted lines).

\subsection{Emergence of the KS relation}
\label{sec:emergence_altfit}

\begin{figure}
\centering\includegraphics{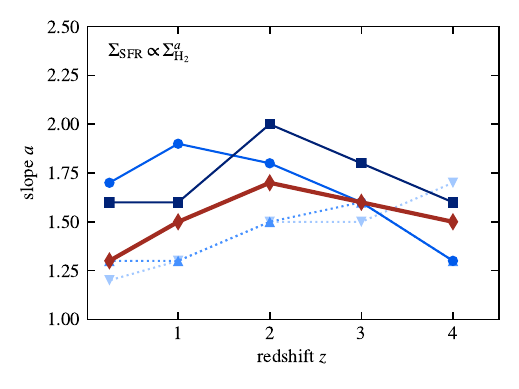}\\
\includegraphics{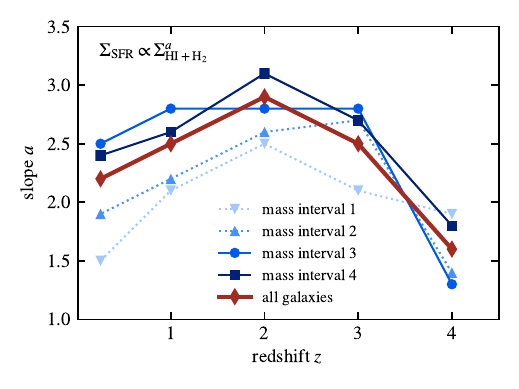}
\caption{Same as Fig.~\ref{fig:emergence}, but using the OLS bisector fitting method (values from Figs.~\ref{fig:KS_SFR10_gasSF_bisec_ssfr} and \ref{fig:KS_SFR10_gastot_bisec_ssfr}).}
\label{fig:emergence_OLS}
\end{figure}

\begin{figure}
\centering\includegraphics{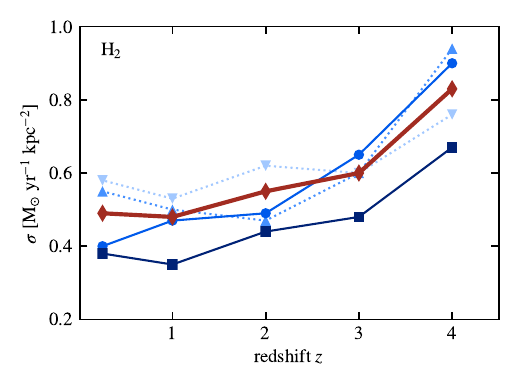}\\
\includegraphics{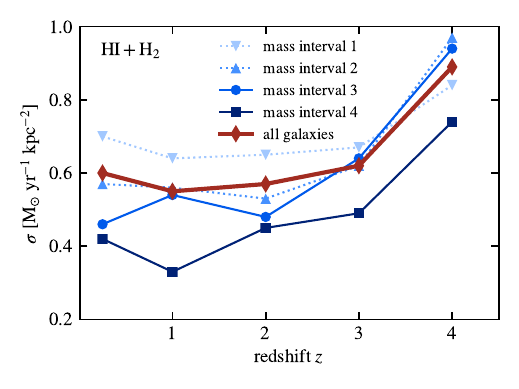}
\caption{Same as Fig.~\ref{fig:emergence_disp}, but using the OLS bisector fitting method (values from Fig.~~\ref{fig:KS_SFR10_gasSF_bisec_ssfr} and \ref{fig:KS_SFR10_gastot_bisec_ssfr}).}
\label{fig:emergence_disp_OLS}
\end{figure}

Figures~\ref{fig:emergence_OLS} and \ref{fig:emergence_disp_OLS}
are the same as Figs.~\ref{fig:emergence} and \ref{fig:emergence_disp}, but for the slopes and dispersions around the best-fit relation using the OLS technique.

\bigskip

To conclude, our analysis reveals that the choice of the fitting method impacts the {\it quantitative} conclusions for all the measurements and diagnostics presented, but that the overall {\it qualitative} trends hold.

\section{Deprojected KS relation}

\begin{figure*}
\centering\includegraphics[width=0.98\textwidth]{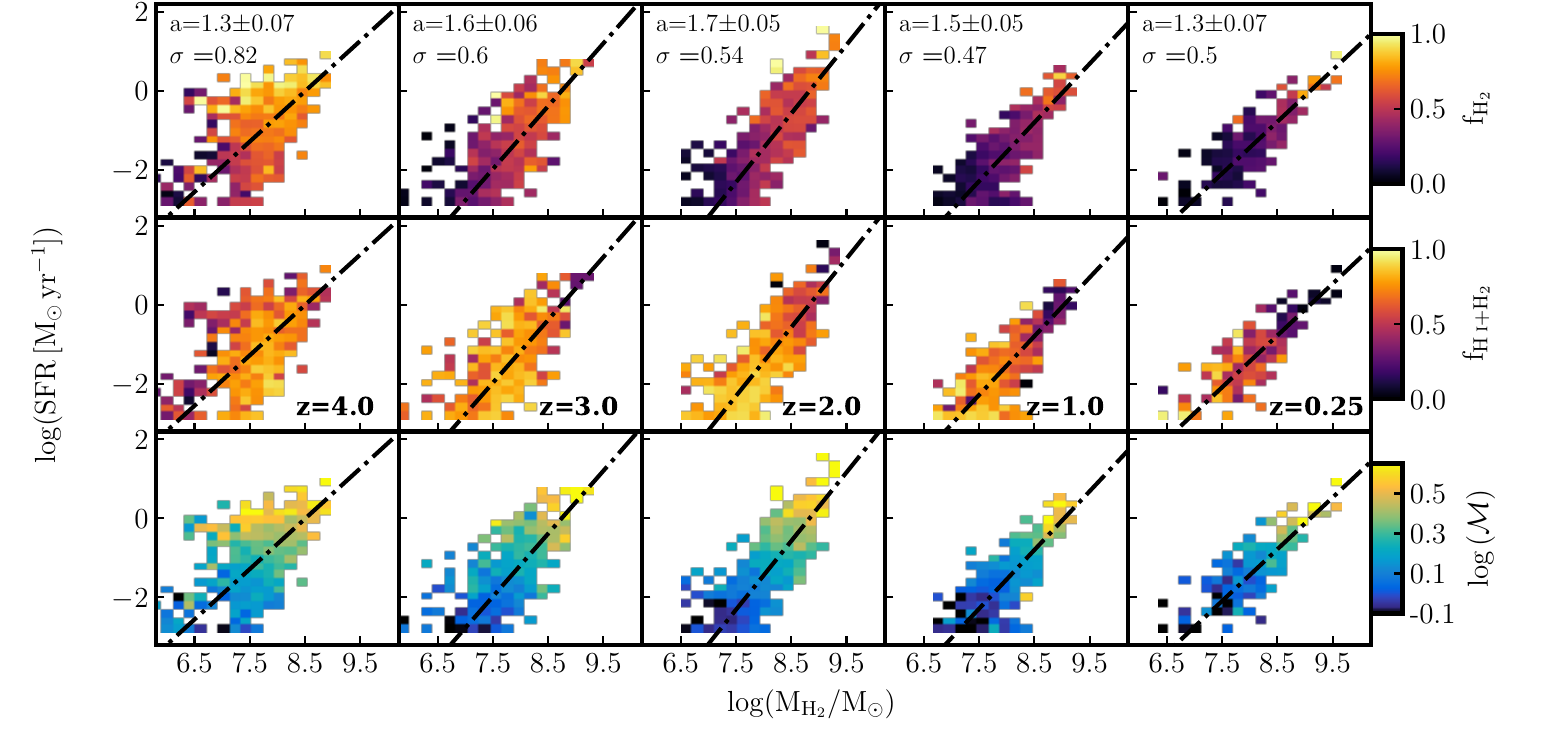}
\caption{Same as \protect Fig.~\ref{fig:KS_int_SFR10_gasSF_reff}, but colour-coded by the molecular gas mass over the total neutral mass (top row), i.e. M$_{\rm{H}_2}$/(M$_{\textsc{H\,i}} + \rm{M}_{\rm{H}_2}$), the neutral baryonic gas fraction (middle row), i.e. (M$_{\rm \textsc{H\,i}} + \rm{M}_{\rm{H}_2})/(\rm{M}_{\textsc{H\,i}} + \rm{M}_{\rm{H}_2} + \rm{M}_\star)$, and the Mach number (bottom row). For comparison, dotted-dash lines correspond to the OLS bisector fits. The corresponding best-fit values for slope $a$ and the standard deviation of residuals $\sigma$ are shown in the top left corners of each panel.
}
\label{fig:KS_int_SFR10_gasSF_3x5}
\end{figure*}

Figure~\ref{fig:KS_int_SFR10_gasSF_3x5} displays the dependence of molecular gas fraction(top row), the neutral gas fraction (middle row), and Mach number (bottom row) of galaxies at different redshifts (columns) on the \mgassf-\sfr plane. The trends seen for the projected quantities \sigmagasSF and \sigmasfr are reproduced, i.e., the only physical parameter that evolves with \sigmagasSF \textit{and} \sigmasfr is the Mach number \mach. 

We find that the results identified in the projected, i.e. observable, versions of the KS plane are also present in their deprojected counterparts, which means that our conclusions are not affected by projection artefacts.

\end{appendix}

\end{document}